\documentclass[aps,prb,reprint,groupedaddress,superscriptaddress,floatfix]{revtex4-2} %Alex: i changed "revtex4-2" to "revtex4-1", because my "to appear" didn't work

\usepackage[T1]{fontenc}
\usepackage{newtxtext}
\usepackage{newtxmath}

\usepackage{amsmath,tikz,calc}
\usepackage{mathtools}
\usepackage{bm}
\usepackage{xfrac}
\usepackage{braket}

\usepackage{color}
\usepackage{xcolor}
\usepackage[colorlinks,
            citecolor=blue,
            linkcolor=blue,
            urlcolor=blue]{hyperref}

\usepackage{graphicx}
\usepackage[all]{hypcap} % Fix reference to figures
\usepackage{lipsum}% http://ctan.org/pkg/lipsum

\usepackage{layout}
\usepackage{bm}   
\usepackage{bbold}

\newcommand{\NS}[1]{{\color{red}{\textbf{NS}:} #1}}

\usepackage{amsmath}

\usepackage{xcolor}
\usepackage{stackengine}
\usepackage{tikz}
\usepackage{mwe}

\usepackage{mpscommands}
\def \tr{{\mbox{tr~}}}

\def \be{\begin{equation}}
\def \ee{\end{equation}}
\def \bea{\begin{eqnarray}}
\def \eea{\end{eqnarray}}

\def\dual#1{\expandafter\dual@aux#1\@nil}
\def\dual@aux#1/#2\@nil{\begin{tabular}{@{}c@{}}#1\\#2\end{tabular}}

\newcommand{\prlsection}[1]{\noindent\textbf{\textit{#1}}--}

\begin{document}

\title{Universality of critical dynamics with finite entanglement}

\author{N. E. Sherman}
\thanks{These two authors contributed equally.}
\address{Department of Physics, University of California, Berkeley, California 94720, USA}
\address{Materials Sciences Division, Lawrence Berkeley National Laboratory, Berkeley, California 94720, USA}

\author{A. Avdoshkin}\thanks{These two authors contributed equally.}
\address{Department of Physics, University of California, Berkeley, California 94720, USA}

\author{J. E. Moore}
\address{Department of Physics, University of California, Berkeley, California 94720, USA}
\address{Materials Sciences Division, Lawrence Berkeley National Laboratory, Berkeley, California 94720, USA}
%\affiliation{Quantum Science Center, Oak Ridge National Laboratory, TN 37831, USA}

\begin{abstract}
When a system is swept through a quantum critical point, the quantum Kibble-Zurek mechanism makes universal predictions for quantities such as the number and energy of excitations produced.  This mechanism is now being used to obtain critical exponents on emerging quantum computers and emulators, which in some cases can be compared to Matrix Product State (MPS) numerical studies.  However, the mechanism is modified when the divergence of entanglement entropy required for a faithful description of many quantum critical points is not fully captured by the experiment or classical calculation.  In this work, we study how low-energy dynamics of quantum systems near criticality are modified by finite entanglement, using conformally invariant critical points described approximately by an MPS as an example.  We derive that the effect of finite entanglement on a Kibble-Zurek process is captured by a dimensionless scaling function of the ratio of two length scales, one determined dynamically and one by the entanglement restriction.  Numerically we confirm first that dynamics at finite bond dimension $\chi$ is independent of the algorithm chosen, then obtain scaling collapses for sweeps in the transverse field Ising model and the 3-state Potts model. Our result establishes the precise role played by entanglement in time-dependent critical phenomena and has direct implications for quantum state preparation and classical simulation of quantum states.
\end{abstract}

\maketitle

% {\hypersetup{linkcolor=black}\tableofcontents}
% {\it Introduction:}
\prlsection{Introduction}
 The long-time dynamics of a many-body quantum system is challenging to study on classical computers even if the system is initialized in a weakly entangled state, as the entanglement entropy will generically grow linearly in time \cite{Calabrese_2005, De_Chiara_2006, PhysRevLett.111.127205, PhysRevB.95.094302}. At the same time, this regime of dynamically produced entanglement is of great interest in modern research, as it contains insights into such fundamental questions as how apparently non-reversible thermalization emerges from unitary dynamics in isolated quantum systems~\cite{Rigol_2008, DAlessio_2016}.  %Questions related to dynamics were also brought to the forefront by the recent advances in material science, where cleaner samples allowed for observation of quasi-1D hydrodynamic behavior \cite{Scheie_2021}, and AMO experiments \cite{Bernien_2017, Zhang_2017}, where the physics of engineered many-body systems can now be probed. \joel{maybe later}
The dynamical aspect is particularly important in quantum simulation on current quantum computers, on which preparing a nontrivial ground state is often harder than performing coherent evolution. Yet, compared to static properties, non-equilibrium time evolution is less understood in terms of either conceptual guiding principles or effective methods of calculation.

An exception is the dynamics of a system swept slowly through a quantum critical point, when universal properties are known to emerge in the limit of long times and distances via the quantum Kibble-Zurek (KZ) mechanism.  We focus on this mechanism as an example of universal out-of-equilibrium dynamics that is  theoretically fundamental and also used in experiments to probe quantum criticality in emerging platforms that maintain quantum coherence well but have difficulty in reaching thermal equilibrium \cite{ebadi2021}.  The modification of quantum criticality by limits on observation time or system size is of renewed interest in light of these new efforts to study such criticality on quantum computers and emulators.  Another, more challenging, kind of modification arises from noise or other effects in the system that act to limit quantum entanglement.  The goal of the present work is to capture how the quantum Kibble-Zurek mechanism is universally modified in systems with finite entanglement.

Quantum critical points are of particular interest because of their emergent universal properties: 
%One of such principles in statistical physics is universality. It is one of the cornerstones in our understanding of complex systems, and according to it,
their large-scale behavior is insensitive to some ``irrelevant'' microscopic details and is the same across vast groups of models known as universality classes.  However, certain other microscopic perturbations are ``relevant'' and change the universality class, and indeed finite entanglement will turn out to be such a perturbation.  Despite having a degree of robustness to irrelevant perturbations, quantum critical points are also well known to be challenging for computational methods on classical computers, for reasons such as requiring large system sizes that also apply to new efforts on quantum computers.  Indeed, finite size can be viewed as a relevant perturbation to criticality, and this insight underlies the successful theory of finite-size scaling \cite{cardyfss}.

%This greatly simplifies the task of describing the long range behavior of a given system by reducing it to the determination of a few parameters defining the universality class. This approach has been particularly useful in the study of phase transitions, where the corresponding parameters are critical exponents capturing the divergences of various observables near the transition point.

Dynamically, the most straightforward manifestation of universality is the (classical or quantum) Kibble-Zurek scaling. It describes the number and energy of excitations produced in a system that is driven through a second-order phase transition. The scaling of the corresponding density with the drive rate is determined by combinations of standard critical exponents. This behavior is often one of the first phenomena probed on new quantum simulation platforms \cite{Bernien_2017, Zhang_2017, PhysRevB.106.L041109}, which has also motivated numerical studies of this process \cite{rams2019symmetry,10.21468/SciPostPhys.9.4.055}.  We derive forms for the fidelity and excitation energy produced by the sweep based on the existence of two relevant scales: the KZ length $\xi_{KZ}$ arising from falling out of adiabaticity with a nonzero sweep rate, and one $\xi_\chi$ arising from the restricted entanglement.

We test the resulting theory using an example of entanglement restriction that is familiar on classical computers: restriction of the bond dimension of a tensor network.  The emergence of this length scale $\xi_\chi$ is a widely used tool in understanding calculations based on Matrix Product States (MPS), and as these calculations are among the most used to model the experimental platforms above, we review their use briefly.

%As such calculations also are widely used to compare to experiments such as those above, we now review the essentials of the MPS approach.  We expect other perturbations inducing finite entanglement to have similar consequences; the advanta
%The numerical approaches most relevant to the experiments are based on Matrix Product States (MPSs) and, in this work, we study the interplay between the limitations of using MPSs and criticality.

MPSs originally emerged as the output of the density matrix renormalization group (DMRG) algorithm \cite{White1992, Schollwock2011}, which provides an approximation to the ground states of 1D local Hamiltonians. The efficiency of this algorithm in many cases is underpinned by the area law of entanglement entropy in gapped one-dimensional systems \cite{Hastings_2007} which implies that the exact ground state can be represented efficiently by an MPS \cite{PhysRevB.73.094423, PhysRevA.75.042306}. Remarkably, for systems with translational symmetry, MPS can represent the exact ground state with a finite number of parameters, even in the thermodynamic limit \cite{PhysRevLett.98.070201}.  Later, MPSs inspired the development of other tensor networks, including the multiscale entanglement renormalization ansatz (MERA) for critical states \cite{PhysRevLett.99.220405, PhysRevLett.101.110501} and projected entanglement pair states (PEPS)  \cite{Verstraete2004, ORUS2014117, Cirac2019, Cirac2021}. MPS applications extend beyond ground state properties to include excited states \cite{PhysRevLett.75.3537, PhysRevB.55.2164, PhysRevB.85.100408, PhysRevLett.111.080401, PhysRevB.97.235155, PhysRevB.88.075133} and quantum dynamics \cite{PhysRevLett.107.070601, PhysRevB.88.075133, PhysRevB.94.165116, Vanderstraeten2019, PAECKEL2019167998, PhysRevB.91.165112, Daley_2004}.

MPSs have also found applications beyond classical simulations of quantum systems. There is a direct mapping between an MPS and certain quantum circuits \cite{PhysRevLett.95.110503, foss2021holographic, haghshenas2022variational}. In such mappings the physical qubits are coupled to some $\chi$-dimensional ancillary system, such as an optical cavity \cite{PhysRevLett.95.110503}, or other qubits \cite{foss2021holographic, haghshenas2022variational}. In this mapping, $\chi$ plays the role of the bond dimension, making it a physically relevant quantity. Recent work has also demonstrated a mapping between tensor networks and neural networks, the main architecture for machine learning (ML) and artificial intelligence \cite{PhysRevB.97.085104, PhysRevLett.122.065301}, allowing for deep learning architectures to be understood from an entanglement perspective \cite{Levine2017_arxiv}. Tensor networks have been successfully used for ML applications, such as image classification \cite{PhysRevB.99.155131, Liu_2019, Efthymiou_2019_arxiv, PhysRevResearch.3.023010, PhysRevX.8.031012, PhysRevB.103.125117, PhysRevResearch.4.043111}. Finally, their property of limiting the total amount of entanglement in the system has also been used as a conceptual tool in the study of bulk reconstruction in the AdS/CFT correspondence \cite{MPSAdS}. %The vast applicability of tensor networks has made them the focus of modern research, rather than just a tool.

For a periodic (or infinite) MPS (iMPS), the expressive power of the ansatz is fully specified by the dimension of the matrices $\chi$, called the bond dimension, which is related to the entanglement entropy of the state \cite{Schollwock2011}; for definitions and details, see Appendix \ref{app:mps}.  However, if the entanglement is unbounded, the existence of an efficient representation of the state with finite $\chi$ is no longer guaranteed. This applies whether the MPS is approximating a critical ground state or the entanglement was dynamically generated.  Here we consider dynamics where, as in many practical computations, the state is represented by an MPS at finite $\chi$ during the full time evolution. We focus on dynamics near a quantum critical point and the goal is to gain insight about properties of the time-evolved state; using the universality of critical behavior, we are able to predict how observables scale with $\chi$ and controllably approach the $\chi=\infty$ state from finite-$\chi$ data.

We begin with a time evolution protocol known as a Kibble-Zurek sweep \cite{Kibble1976, Kibble1980,zurek1985cosmological, Zurek2014}. We then
%address the subtleties of what is meant by finite $\chi$ dynamics, as well as the finite $\chi$ ground state. In particular, we
show that finite $\chi$ dynamics is well-defined, in that different procedures for time evolution produce the same result in the appropriate limits. A subtlety is that different definitions that all give the exact ground state are no longer equivalent at finite $\chi$, and how algorithms resolve this ambiguity.  We then demonstrate our results by examining the transverse-field Ising model (TFIM) and the 3-state Potts model, verifying our finite $\chi$ scaling hypothesis in detail. %Lastly, we conclude with a summary, and consequences of this work.

\prlsection{Kibble-Zurek Scaling}
We consider an extended quantum system described by a Hamiltonian $H(\lambda)$ with some parameter $\lambda$. We further assume that $\lambda = \lambda_0$ corresponds to an isolated quantum critical point. For  the correlation length $\xi$ and time $\tau$ in the vicinity of the critical point we expect \cite{sachdev1999quantum}

\bea \label{critical_exps}
\xi \sim |\lambda-\lambda_0|^{-\nu}, \tau \sim |\lambda-\lambda_0|^{-z \nu},
\eea
where $\nu$ and $z$ are the corresponding critical exponents.

Let us now consider the evolution of the system initiated in the ground state (that we assume to be non-degenerate) far away from the critical point with the parameter changing in time as $\lambda(t) = \lambda_0 + \varv t$. We assume that $v$ is slow compared to the bandwidth and $t$ runs from $-\infty$ to $t_0>0$. Far from the critical point the gap is large compared to $v$ and the adiabatic theorem applies. Because the breakdown of adiabaticity only occurs close to the critical point, properties of the resulting state will obey universal scaling laws, the KZ scaling \cite{Kibble1976, Kibble1980,zurek1985cosmological, Zurek2014}.

The scaling exponents can be deduced from a simple reasoning. The adiabaticity is lost when $t \approx -\tau$, where $\tau$ is determined by Eq. (\ref{critical_exps}); this corresponds to
\bea
\tau_{KZ} \sim \varv^{-\frac{\nu z}{1 + \nu z}},~ \xi_{KZ} \sim \varv^{-\frac{\nu}{1 + \nu z}},
\eea
thus defining the Kibble-Zurek time and length, correspondingly.
Since the adiabaticity is restored after $t = \tau$ and we expect the generated exitations to freeze out and the average density of excitations and energy will be (in one spatial dimension)
\bea \label{eq:n_KZ}
n_{ex} \sim 1/\xi_{KZ} \sim \varv^{\frac{\nu}{1 + z \nu}},\\
\epsilon_{ex} \sim 1/\xi^2_{KZ} \sim \varv^{\frac{2\nu}{1 + z \nu}}.
\eea
This scaling has been verified by extensive numerics \cite{DeGrandi2010, ROSSINI20211} as well as experiments \cite{Clark_2016, Bernien_2017, Zhang_2017}. There exists also an exact solution for the transverse-field Ising model \cite{dziarmaga2005dynamics, PhysRevA.73.043614}.

In Eq. \eqref{eq:n_KZ}, $\epsilon_{ex}$ is the energy above the ground state divided by the volume and $n_{ex}$ needs to be defined with care when particle number is not well-defined. We propose to use fidelity density, which is given by
\be\label{eq:fidelity}
f(t) = - \frac{1}{N} \log \left(|\langle \psi(t) | \psi_0 \rangle|^2\right),
\ee
where $| \psi_0 \rangle$ is the ground state, $\ket{\psi(t)}$ the time evolved state, and $N$ is the total number of sites in the system. $f(t)$ has the same scaling as we expect for $n_{ex}$ %\cite{PhysRevB.81.224301}
\cite{PhysRevB.84.224303,DeGrandi2010}, and is proportional to it at low densities when the system has a free fermion description, as we explain in Appendix \ref{app:fidelity}.

\prlsection{MPS Dynamics and Finite Bond Dimension}
The time evolution of a state under a Hamiltonian $H(t)$ is given by
\begin{align}
    \ket{\psi(t)} &= U(t) \ket{\psi(0)} \\
    U(T) &= \mathcal{T}\exp\left(-i\int_0^Tdt\, H(t)\right)
\end{align}
where $\mathcal{T}(\cdot)$ is the time ordering operator. If we write $t_n=n\Delta t$, and $T=t_N$, then this is equivalent to writing
\begin{equation}
  U(t) = \lim_{N\rightarrow \infty}\left[e^{-iH(t_N)\Delta t} \cdots e^{-iH(t_0)\Delta t} \right].
\end{equation}
Conceptually, this amounts to treating the Hamiltonian as piecewise constant over an interval of size $\Delta t$, and the exact time evolution is found in the limit that $\Delta t\rightarrow 0$. For finite $\Delta t$, treating the Hamiltonian as piecewise constant produces an error of order $\mathcal{O}(\Delta t)$. To implement time evolution using an MPS, if $\Delta t$ is sufficiently small, it is sufficient to define time evolution for a constant Hamiltonian over a time $\Delta t$.

In practice, the time evolution of a state is implemented in one of three ways. Most commonly used when studying a nearest-neighbor Hamiltonian is the Time Evolution Block Decimation (TEBD) algorithm \cite{Daley_2004, Schollwock2011}, which is based on a Suzuki-Trotter decomposition of the time evolution operator. One can also calculate a matrix product operator (MPO) representation of the matrix exponential itself \cite{PhysRevB.91.165112}. For both of these methods, a time evolution step will generically increase the entanglement of the system, and thus increase the bond dimension $\chi$. Thus a procedure is required to project the resulting state down to a state with a bond dimension $\chi_{\rm{max}}$. The last notable time evolution procedure, the time-dependent variational principle (TDVP) \cite{PhysRevLett.107.070601, PhysRevB.88.075133, PhysRevB.94.165116, Vanderstraeten2019}, finds the state in the manifold of $\chi \le \chi_{\rm{max}}$ that minimizes the $l_2$ difference with the exact time evolved state.  Often we are interested in the limit of both $\chi\rightarrow \infty$ and $\Delta t\rightarrow 0$, corresponding to the exact solution of the time-dependent Schrodinger equation. In this work, we hold $\chi$ finite and discuss just the limit $\Delta t\rightarrow 0$. It is the case that these three methods of time evolution produce the same state when $\chi$ is fixed and $\Delta t \rightarrow 0$; see Appendix \ref{app:equiv}.

Now we turn to how KZ scaling is modified when the state is represented at all times by an MPS with a fixed finite bond dimension $\chi$. The effect of finite bond dimension is to limit the amount of entanglement in the system. And since close to conformal critical points entanglement is related to the correlation length by the celebrated expression \cite{Calabrese_2009}
\bea
S = \frac{c}{6} \log \xi,
\eea
for fast sweeps, when $\xi_{KZ}$ is small, the effect of $\chi$ will be small, whereas for slower sweeps, when $\xi_{KZ}$ is large, the number of excitations will be suppressed compared to Eq. \eqref{eq:n_KZ}.

This situation is similar to the one studied in \cite{rams2019symmetry}, where the KZ scaling in the TFIM was studied in the presence of a symmetry breaking bias $g_{||}$ that kept the gap finite at all times during the sweep. It was numerically verified that the effect of $g_{||}$ could be described by a single length scale $\xi_{||} = g_{||}^{-\nu_{||}}$ where $\nu_{||}$ is the corresponding critical exponent, and all KZ scaling laws were modified by scaling function that depended on the ratio $\xi_{||}/\xi_{KZ}$. We also expect the scaling behavior to occur for finite system size, in which case the argument of the scaling function would be $L/\xi_{KZ}$ with $L$, the system size.

Returning to the case of finite bond dimension, we conjecture its effect to be describable by a single length scale $\xi_{\chi}$. Thus, we expect that the $\chi=\infty$ result is modulated by a dimensionless scaling function, similar to the scaling theory of entanglement entropy \cite{PhysRevB.98.245124}. In particular, we expect
\bea\label{eq:n-chi}
\mathcal{O}(\varv,\chi) = \mathcal{O}(\varv, \chi=\infty) f_{\mathcal{O}}(\xi_{KZ} / \xi_{\chi})
\eea
% \bea\label{eq:n-chi}
% n_{ex} = v^{\frac{\nu}{1 + \nu}} f(\xi_{KZ}/\xi_{\chi}),\quad \epsilon = v^{\frac{2\nu}{1 + \nu}} g(\xi_{KZ}/\xi_{\chi}),
% \eea
where $f_{\mathcal{O}}$ is some scaling function for the observable $\mathcal{O}$. Here, we look at the fidelity density $f$ from Eq.~(\ref{eq:fidelity}) and the excitation energy $\epsilon_{ex}$. the fidelity density $f$ is computed via the largest eigenvalue of the transfer matrix formed by the full contraction of both states (Appendix \ref{app:fidelity}).

% It is important to note that, since, both $n_{ex}$ and $\epsilon_{ex}$ are defined with respect to the ground state we should use the finite-$\chi$ ground state, i.e. the MPS with bond dimension $\chi$ that has the lowest expectation of $H$ - the output of the DMRG algorithm.

The length scale $\xi_{\chi}$ has been previously studied for ground state properties and the scaling given by $\xi_{\chi} \sim \chi^{\kappa}$ was observed in \cite{tagliacozzo2008scaling}. The conformal field theory (CFT) entanglement spectrum was used to obtain a form for the exponent \cite{pollmann2009theory}
\bea\label{eq:kappa}
\kappa = \frac{6}{c \left( \sqrt{\frac{12}{c}} + 1\right)}
\eea
in surprisingly good agreement with numerical data \cite{tagliacozzo2008scaling,pollmann2009theory,pirvu2012}. Looking ahead, we will find that the same critical exponent governs the dynamical problem of Kibble-Zurek scaling.

To calculate the fidelity density $f$ and the excitation energy $\epsilon_{\mathrm {ex}}$, the ground state is also needed. Defining the correct ground state is subtle and requires some care. The ideal state to compare the time-evolved state against is the adiabatic state
\begin{equation}\label{eq:slow_ground_state}
    \ket{\psi_0} := \lim_{\varv\rightarrow 0}\ket{\psi(\varv,t=0)}
\end{equation}
Analytically, at infinite $\chi$, this definition is equivalent to the exact ground state of the Hamiltonian. However, at finite $\chi$, different symmetry requirements on the ground state can produce different states. For example, if the Hamiltonian has a symmetry, and the initial state is an eigenstate of the symmetry operator, then the time-evolved state will also be symmetric. However, if the ground state is degenerate, the symmetry-broken state at fixed $\chi$ has a lower energy, and will be found by a generic ground state search. To address this, we compute the ground state using the same parameters and starting state as the time evolution process, but perform a symmetry-preserving imaginary time evolution to cool into the ground state.

\prlsection{Numerical Verification}
We look at two models in this study. First, the transverse-field Ising model (TFIM), defined by the Hamiltonian
\bea \label{eq:IsingHam}
H = -J\sum_n\sigma_n^z\sigma_{n+1}^z - g\sum_n\sigma_n^x,
\eea
where $\sigma^{i}_n$ is the $i$-th Pauli matrix at site $n$. Eq. \eqref{eq:IsingHam} has a $Z_2$ symmetry $\bigotimes\limits_i \sigma^x_i$.  This system has a quantum phase transition at $g = J$ that separates a disordered phase with a unique GS for $g>J$ and an ordered phase with a two-fold degenerate GS for $0<g<J$. The CFT describing the critical point is the minimal model with $c = 1/2$ and the critical correlation length critical exponent is given by $\nu = 1$ \cite{francesco2012conformal}. Thus, for the KZ scaling we expect
\bea
n_{ex} \sim \varv^{1/2},~ \epsilon_{ex} \sim \varv.
\eea

The coupling constants have a $\varv$ dependence given by
\begin{align}\label{eq:coupling-time}
    \begin{matrix}
    J(t) = 1 + \varv t \\
    g(t) = 1 - \varv t 
    \end{matrix}\quad
    t:-\frac{1}{\varv}\rightarrow 0
\end{align}
The initial coupling is given by $J=0$, and $g=2$. The ground state at this point is a simple product state given by $\ket{\psi_0} = \ket{\rightarrow}^{\otimes N}$. We then time evolve this state with the time dependent Hamiltonian using the time-evolution block-decimation (TEBD) algorithm \cite{Daley_2004, Schollwock2011}. We use a fourth order Trotter decomposition \cite{Barthel_2020}, with a timestep of $dt=0.005$. 

\begin{figure}
    \centering
    \includegraphics[width=\linewidth]{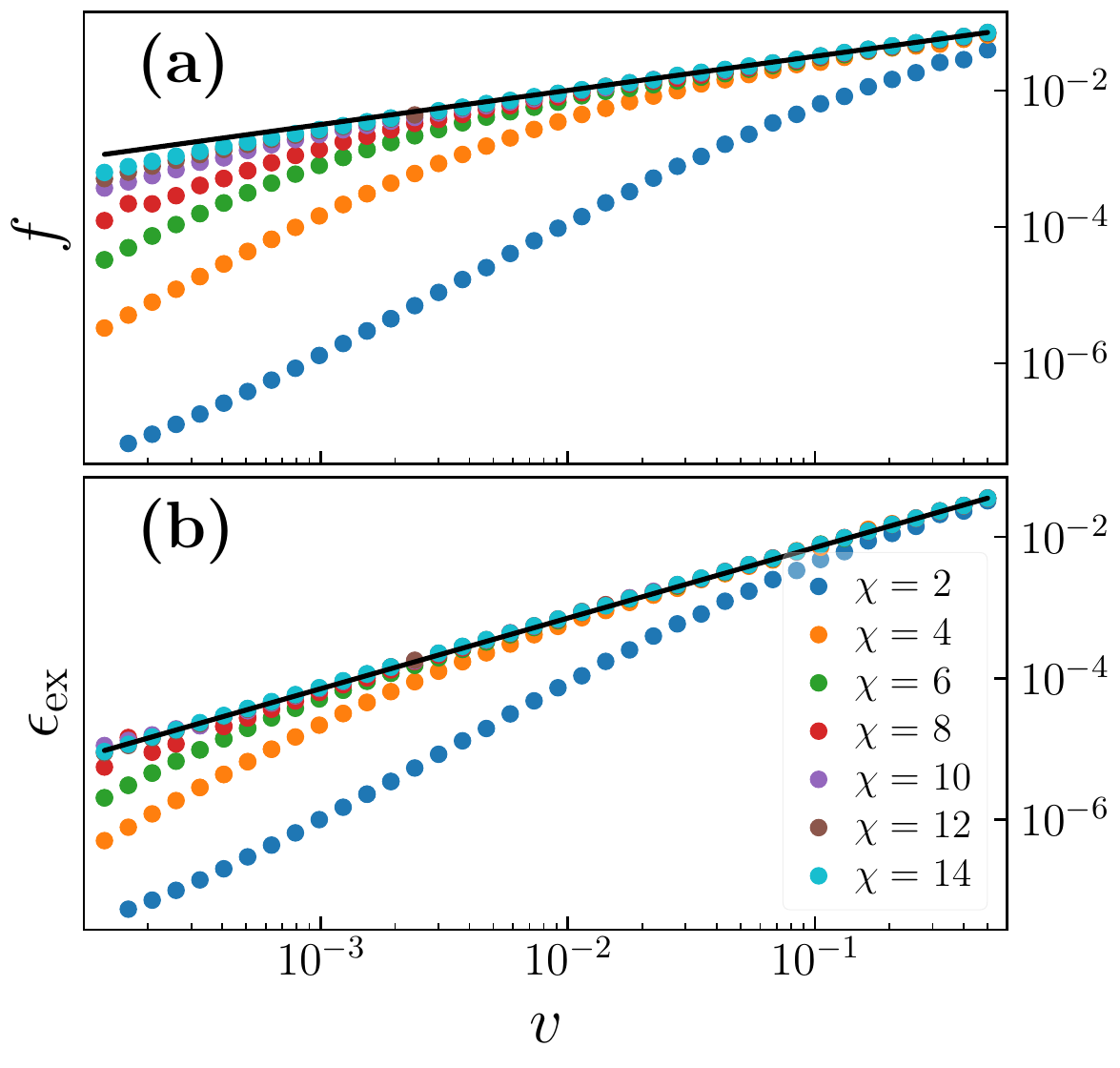}
    \caption{The fidelity and excitation energy densities after a Kibble Zurek sweep performed at speed $v$ for the TFIM. We show the results for different maximum bond dimensions $\chi$. We show a black line illustrating the scaling prediction for $\chi=\infty$.}
    \label{fig:Ising_Ev_Nv}
\end{figure}

We enforce the $\mathbb{Z}_2$ symmetry during the ground state search, and time evolution, producing a $\mathbb{Z}_2$ symmetric state in both cases. For different values of the bond dimension $\chi$, we calculate $f$ and $\epsilon_{\mathrm{ex}}$, and show the results in Fig. \ref{fig:Ising_Ev_Nv}. The black line illustrates the $\chi=\infty$ result. We see that for large $\varv$, the effect of finite bond dimension is minimal, but as we decrease the speed, the deviations become dramatic. The systematic nature of the deviations is a focus of the present work.

\begin{figure}
    \centering
    \includegraphics[width=\linewidth]{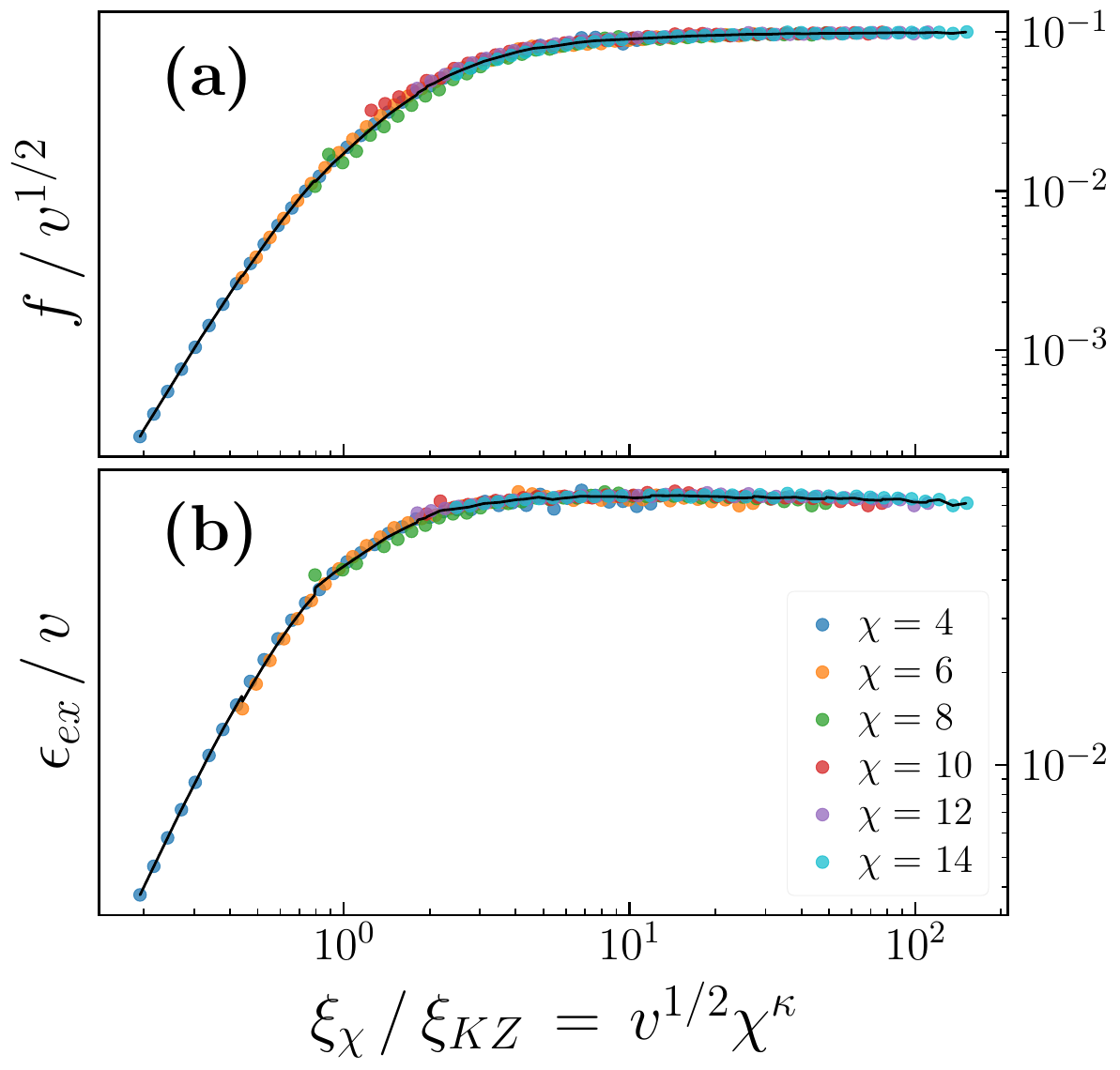}
    \caption{The scaling function collapse for the fidelity density and excitation energy in the TFIM. The length scale introduced by the bond dimension, $\xi_{\chi}$, follows a power law with exponent given by Eq. \eqref{eq:kappa}, with a central charge of $c=1/2$.}
    \label{fig:Ising_scaling}
\end{figure}

In Fig. \ref{fig:Ising_scaling}, we show the scaling function collapse assuming the scaling hypothesis in Eq. \eqref{eq:n-chi}. We expect the scaling hypothesis to be valid for large $\chi$ but it already begins to work for $\chi \geq 4,$ with the exponent $\kappa$ specified by Eq. \ref{eq:kappa}.

The second model we explore is the 3-state Potts model defined by the Hamiltonian \cite{Solyom_1981}
\begin{align}
    H = -J\sum_n \left(\eta_n^{\dagger}\eta_{n+1}+\eta_{n}\eta_{n+1}^{\dagger}\right)
        &-g\sum_n\left(\tau_n + \tau_n^{\dagger}\right)\label{eq:Potts_quantum_ham}\\
    \eta= \begin{bmatrix}
        1 & &  \\
        & e^{i\frac{2 \pi}{3}} &  \\
        & & e^{-i\frac{2 \pi}{3}}
    \end{bmatrix},~~&
    \tau= \begin{bmatrix}
        0 & 1 & 0 \\
        0 & 0 & 1 \\
        1 & 0 & 0
    \end{bmatrix}
\end{align}
    
Eq. \eqref{eq:Potts_quantum_ham} enjoys $S_3$ symmetry comprised of all permutation of the basis states on all sites\iffalse , generated by $\bigotimes\limits_i \tau_i$\fi. However, we explicitly enforce the only $\mathbb{Z}_3 \subset S_3 $ symmetry containing cyclic permutations.

Analogously to the TFIM, at $g=J$ there is a critical point that separates a $Z_3$ symmetric phase for $g>J$ and a $Z_3$ ordered phase for $0<g<J$. The CFT is similarly a minimal model ($Z_3$ parafermion) with $c = 4/5$ and $\nu = 5/6$ \cite{francesco2012conformal}. Accordingly, the KZ scaling is

\bea
n_{ex} \sim \varv^{5/11},~~ \epsilon_{ex} \sim \varv^{10/11}.
\eea

\begin{figure}
    \centering
    \includegraphics[width=\linewidth]{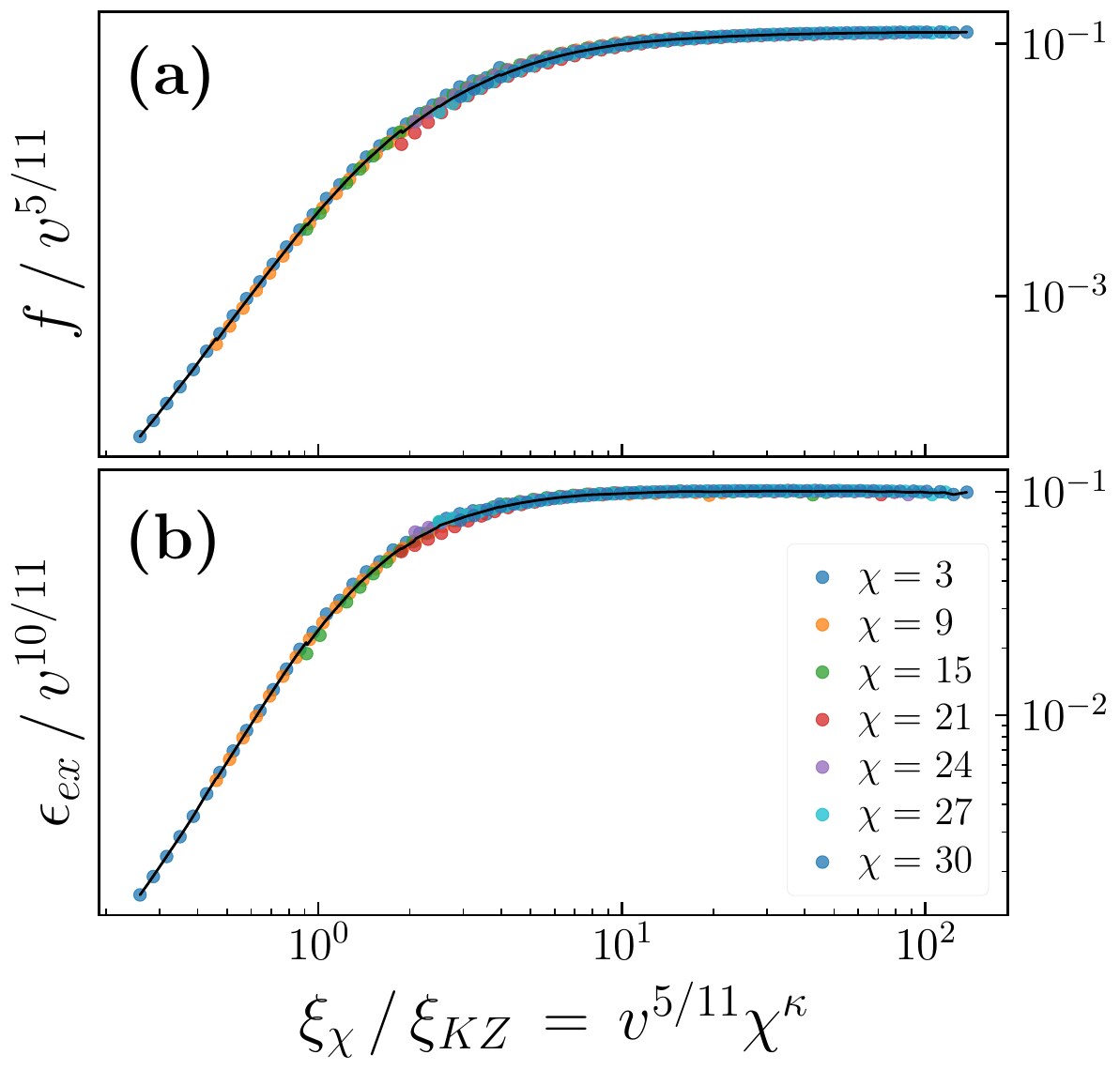}
    \caption{The scaling function collapse for the fidelity density and excitation energy in the 3-state Potts model. The length scale introduced by the bond dimension, $\xi_{\chi}$ is a power law with exponent given by Eq. \eqref{eq:kappa}, with a central charge of $c=4/5$.}
    \label{fig:Potts_scaling}
\end{figure}

For the dynamics, we use the same time dependent coupling used for the TFIM, given in Eq. \eqref{eq:coupling-time}. We again use TEBD with a fourth order Trotter decomposition, except with a timestep of $dt=0.01$. The excitation energy, and fidelity density, are qualitatively identical to the TFIM, except with different scaling exponents with $\varv$, and so we relegate the figures to Appendix \ref{app:fidelity}. We do show the scaling function collapse in Fig. \ref{fig:Potts_scaling}. Again we see great collapse of the data, further confirming the scaling hypothesis of Eq. \eqref{eq:n-chi}.

% \begin{figure}
%     \centering
%     \includegraphics[width=\linewidth]{Potts_Ev_Nv.pdf}
%     \caption{The fidelity density, and excitation energy after a Kibble Zurek sweep performed at speed $v$ for the 3-state Potts model. We show the results for different maximum bond dimensions $\chi$. We show a black line illustrating the scaling prediction for $\chi=\infty$.}
%     \label{fig:Potts_Ev_Nv}
% \end{figure}

We want to note that we omit the data for $\chi=6, 12, 18$ in the 3-state Potts model. The scaling function collapse is not found for small $\varv$ for these values of $\chi$. We believe this is because of a difference in symmetry between the ground state and the time-evolved state. Since the 3-state Potts model has a full $S_3$ symmetry, the time evolved state will produce a state that is fully $S_3$ symmetric. However, since we only enforce a $\mathbb{Z}_3$ symmetry, the ground state search will break the $S_3$ symmetry down to a $\mathbb{Z}_3$ symmetric state, as this will reduce the energy. Our numerical simulations do not implement non-Abelian quantum numbers, making enforcing the $S_3$ symmetry difficult. We expect that enforcing this symmetry will cause the scaling function to collapse for $\chi$ a multiple of $6$ as well.

\prlsection{Conclusions}
We found that a Kibble-Zurek sweep through a one-dimensional quantum critical point is modified by finite entanglement, i.e., fixed finite bond dimension $\chi$ for an iMPS, in a way similar to relevant perturbations of the Hamiltonian, even though finite $\chi$ is not equivalent to any local Hamiltonian perturbation.  Properly defined, the sweep-induced differences from an adiabatically defined ground state are captured by a universal scaling function that unusually involves both scaling dimensions and central charge.  The scaling function involves the ratio of two length scales $\xi_{KZ}$ and $\xi_{\chi}$ and the essential features are independent of the specific implementation of the dynamics, suggesting that the finite-entanglement scaling form for dynamics will have similar utility in practice as the form for ground states, by enabling systematic extrapolation from finite-$\chi$ results (see Appendix \ref{app:extrapolation} for further details).

Whether bond dimension can be treated as a relevant perturbation in an even more general setting, and whether other non-Hamiltonian perturbations to quantum dynamics can similarly be captured by scaling functions, remain an open questions.  Actual perturbations in quantum hardware could certainly be more relevant than the finite-entanglement restriction, but a natural conjecture is that any local hardware with limits on achieved entanglement (i.e., Schmidt rank) can do no better than the matrix product states studied here.
It would be interesting to see if our analysis applies beyond Kibble-Zurek scaling to the more general finite-time scaling \cite{Gong_2010,PhysRevB.90.134108}. Lastly, the finite $\chi$ scaling of dynamical observables opens up an interesting application of quantum computers in the NISQ era. Since quantum circuits can represent MPSs with a physically relevant bond dimension \cite{PhysRevLett.95.110503, foss2021holographic, haghshenas2022variational}, running such simulations at different bond dimensions could enable a novel way to extract the central charge of critical theories. This procedure is well suited for quantum computers, on which unitary dynamics are easily programmed.

\begin{acknowledgments}
This work was primarily supported by the National Science Foundation through QLCI grant OMA-2016245 (A.A.) and grant DMR-1918065 (J.E.M.).  N.S. received support from the U.S. Department of Energy, Office of Science, Office of Basic Energy Sciences, Materials Sciences and Engineering Division under Award No. DE-AC02-05-CH11231 through the Theory Institute for Materials and Energy Spectroscopy (TIMES). J.E.M. was also supported by a Simons Investigatorship. A. A. acknowledges support from a Kavli ENSI fellowship. This research also used resources of the National Energy Research Scientific Computing Center (NERSC), a U.S. Department of Energy Office of Science User Facility operated under Award No. DE-AC02-05CH11231. 
\end{acknowledgments}

\newpage

\appendix

\begin{widetext}

% \section{MPSs and tangent space methods} 
\section{Matrix Product States} 
\label{app:mps}

MPSs are defined as the following wave-function ansatz

\bea \label{eq:mps_def}
\langle s_1 s_2 \cdots s_{n-1} s_n |\Psi(A)\rangle = \tr \left[A^{(1)}_{s_1} A^{(2)}_{s_2} \cdots A^{(n-1)}_{s_{n-1}} A^{(n)}_{s_n} \right], 
\eea
where $s_i$ runs over the indices of the local Hilbert space and for each value of $s_i$, 
$A^{(i)}_{s_i}$ is a $\chi\times\chi$ matrix. Pictorially, we can represent this state as a tensor network

\begin{equation} 
\ket{\Psi(A)} =  \dots
\begin{diagram}
\draw (0.5,1.5) -- (1,1.5); 
\draw[rounded corners] (1,2) rectangle (2,1);
\draw (1.5,1.5) node (X) {$A$};
\draw (2,1.5) -- (3,1.5); 
\draw[rounded corners] (3,2) rectangle (4,1);
\draw (3.5,1.5) node {$A$};
\draw (4,1.5) -- (5,1.5);
\draw[rounded corners] (5,2) rectangle (6,1);
\draw (5.5,1.5) node {$A$};
\draw (6,1.5) -- (7,1.5); 
\draw[rounded corners] (7,2) rectangle (8,1);
\draw (7.5,1.5) node {$A$};
\draw (2.1,0.75) node {$i-2$};
\draw (4.1,0.75) node {$i-1$};
\draw (5.75,0.75) node {$i$};
\draw (8.1,0.75) node {$i+1$};
\draw (10.1,0.75) node {$i+2$};
\draw (8,1.5) -- (9,1.5); 
\draw[rounded corners] (9,2) rectangle (10,1);
\draw (9.5,1.5) node {$A$};
\draw (10,1.5) -- (10.5,1.5);
\draw (1.5,1) -- (1.5,.5); \draw (3.5,1) -- (3.5,.5); \draw (5.5,1) -- (5.5,.5);
\draw (7.5,1) -- (7.5,.5); \draw (9.5,1) -- (9.5,.5); 
\end{diagram} \dots,
\end{equation}
where the dangling vertical legs correspond to the physical indices of the wavefunction and the horizontal legs represent the matrix multiplication in Eq. \eqref{eq:mps_def}.

% \section{TEBD=TDVP for $dt \to 0$} 
\section{Equivalence Between TEBD and TDVP} 
\label{app:equiv}

Here we will consider the following question. Starting with an MPS $\ket{\Psi(A)}$ with bond dimension $\chi$, we perform one step of time evolution, corresponding to a time step $dt$, that increases the bond dimension to $\chi'$. Finite $\chi'$ can be achieved by considering an arbitrary but finite number of Trotter-Suzuki steps. We denote the new MPS as $\ket{\Psi(A+B)}$, where $B$ has bond dimension $\chi'$, and we can enlarge A with zeros to make the sum $A+B$ well defined. We wish to understand whether the different ways of compressing this state back to bond dimension $\chi$ become equivalent when we take $dt \to 0$. In this limit $B$ can be thought of as a tangent vector \cite{vanderstraeten2019tangent}.

Let us denote the compressed state $| \Psi(A + B')\rangle$ where $B'$ is an MPS of bond dimension $\chi$. The difference between TEBD and WII MPO/TDVP can be thought of as the difference in how truncation is performed. The TEBD projection of $A+B$ to the subspace of bond dimension $\chi$ corresponds to the truncation of the SVD spectrum \cite{vidal2004efficient, PhysRevLett.93.207204}. This would be an optimal compression in terms of vector distance $\left| |\Psi(A+B')\rangle - |\Psi(A+B)\rangle \right|^2$ if we were only changing the tensor from $A+B$ to $A + B'$ at one site \cite{Schollwock2011}, but is not optimal for a global change. Meanwhile for the WII MPO method the globally optimal
$\left| |\Psi(A+B')\rangle - |\Psi(A+B)\rangle \right|^2$ \cite{PhysRevB.91.165112} is achieved and the TDVP is the infinitesimal version of that \cite{vanderstraeten2019tangent}.

Let us now show that both compression are equivalent to the linear order in $B,B'$.
We expand the states to the linear order in B,B', obtaining
\begin{equation} 
\ket{\Psi(A + B)} \approx \ket{\Psi(A)} + \sum_i \dots
\MPSMixed{$A$}{$B$}{$A$} \dots.
\end{equation}

Thus to the first order, the difference of the states to be minimized is

\begin{equation} 
\ket{\Psi(A + B)} - \ket{\Psi(A + B')} \approx \sum_i \dots \MPSMixedBigCenter{$A$}{$B-B'$}{$A$}
 \dots.
\end{equation}

The last step is to contract it with itself. For definiteness work with the left gauge for the tangent vector $B$, in which case

\begin{equation} \label{eq:gaugecondition}
\applyTransferLeft{$\bar{A}$}{$l$}{$B$} = \applyTransferLeft{$\bar{B}$}{$l$}{$A$}= 0 .
\end{equation}

With that gauge, the contraction of the difference with itself will only have diagonal components, where $A$'s are contracted with $A$'s and $B-B'$ is contracted with itself at site $i$:

\begin{align}
\left|\ket{\Psi(A + B)} - \ket{\Psi(A + B')}\right|^2 \approx \sum_i
\dots \MPSFullContractionBigCenter{$A$}{$\bar{A}$}{$B-B'$}{$\bar{B} - \bar{B}'$}{$A$}{$\bar{A}$} \dots \nonumber
\end{align}
This is the same as what we would obtain if we just wanted to truncate a nonuniform MPS
\begin{equation} 
\ket{\Psi(A,B,i)} =  \dots
\MPSMixedBigCenter{$A$}{$A+B$}{$A$}
\dots.
\end{equation}

(no sum over $i$!) at a single site, where we know the truncation of the SVD spectrum is optimal. This proves that both ways of truncation are equivalent when $dt\to 0$ and, thus, TEBD, TDVP and WII MPO are also equivalent in this limit.

We also demonstrate the equivalence between these methods numerically, see Fig. \ref{fig:tebd-tdvp}. We work at $\chi=16$, starting with the ground state of the critical TFIM, i.e. Eq. \eqref{eq:IsingHam} with $(J,g) = (1,1)$. We then time evolve this state using the coupling $(J,g)=(0.1, 1)$, for $t=J$. As $dt$ decreases, we see that all three methods converge.

\begin{figure}
    \centering
    \includegraphics[width=0.5\linewidth]{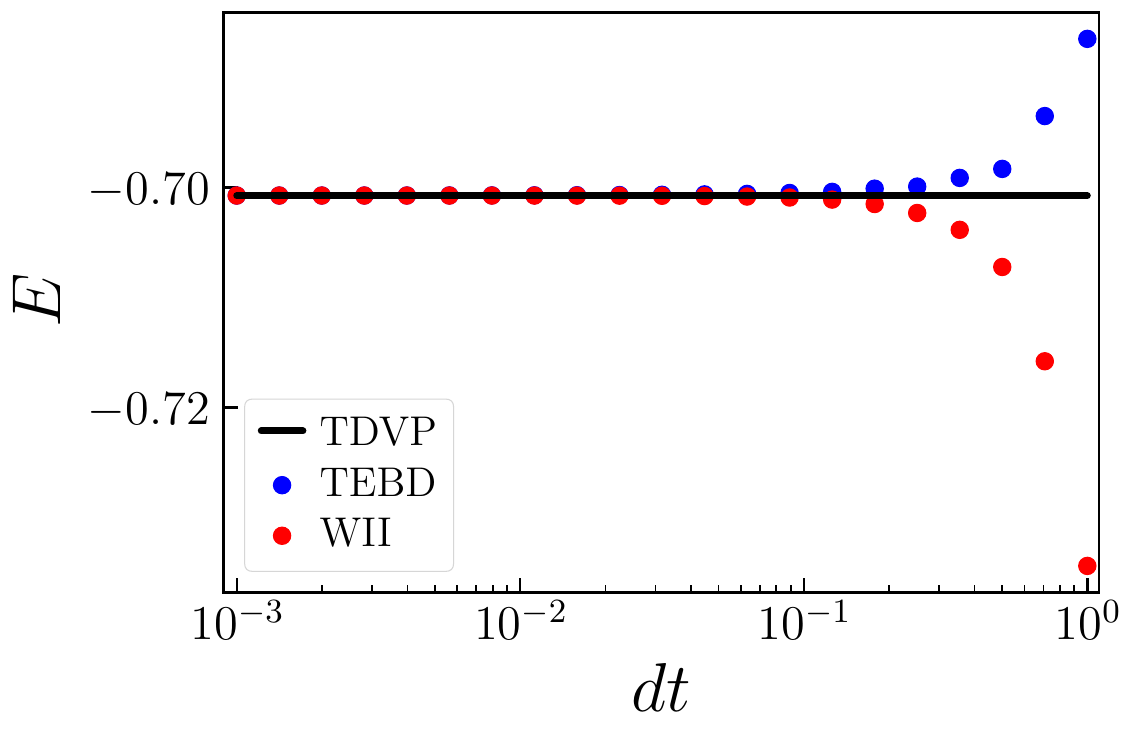}
    \caption{The energy of the time evolved state obtained from TEBD and TDVP, for different values of the timestep $dt$ used in TEBD. To prepare the state, we started with the $\chi=16$ ground state of the critical TFIM $(J,g) = (1,1)$, and then time evolved with the the TFIM with $(J,g) = (0.1,1)$ for $t=J$. The TDVP data was obtained using a timestep of $10^{-3}$, and a 2-site iTDVP algorithm. Maybe cat state issues with other quantities, could try a different time evolution process.}
    \label{fig:tebd-tdvp}
\end{figure}

\section{Fidelity and Excitation Density} 
\label{app:fidelity}

In this section, we will use the exact solution of the KZ sweep for the TRIM \cite{dziarmaga2005dynamics} to show that the fidelity density reduces to the excitation density in the low-density limit. Let $|\psi(t)\rangle$ be the time evolved state after a Kibble-Zurek sweep, and $|{\psi_0}\rangle$ the ground state. Then, using the free fermion representation from \cite{dziarmaga2005dynamics}, the time-evolved state can be written as
\bea
|\psi(t)\rangle = \prod_{k>0} (\alpha_k(t) | \text{no pair} \rangle_k + \beta_k(t) | \text{ pair} \rangle_k),
\eea
where $|\text{pair}\rangle_k$ denotes the state with quasipartice of momenta  $k$ and $-k$ present above the ground state. The probability of finding the $k,-k$ pair in this state is $P_k = |\beta_k(t)|^2$. The expected number of particles in $|\psi(t)\rangle$ is $N_{ex} = 2 \sum\limits_{k>0} P_k$ and for the fidelity we find
\bea\label{eq:purity}
|\langle \psi_{0} | \psi(t) \rangle|^2 = \prod_k|\langle \psi_k^{(0)}|\psi_k(t)\rangle|^2 = \exp \left( \sum_{k>0} \log (1 - P_k)  \right) \approx \exp \left( - \sum_{k>0} P_k  \right) = e^{-N_{ex}/2},
\eea
where we have used that the excitation density is low $P_k \ll 1$ (but not $\sum_{k>0} P_k \ll 1$). 
Now, defining the excitation density as $n_{ex} = N_{ex} / L$, and we find that the fidelity density is given by
\begin{equation}
    f(t) := -\frac{1}{N}\log\left(|\langle \psi_{0} | \psi(t) \rangle|^2\right) \approx n_{ex} / 2
\end{equation}
This approximation becomes exact in the limit of low-density of excitations. In the context of a Kibble-Zurek sweep, this limit corresponds to $\varv\rightarrow 0$.

To compute this quantity using uniform MPSs, we note
\begin{equation}
    \langle \psi_{0} | \psi(t) \rangle = \text{tr}\left(
    \overlapTransferMatrix{$T$}{$L$}\right),
\end{equation}
where
\begin{equation}
    \transferMatrix{$T$}{$A(t)$}{$\bar{A}_0$}.
\end{equation}
If $\lambda$ is the magnitude of the largest eigenvalue of $T$, then in the thermodynamic limit we have
\begin{align}
    |\langle \psi_{0} | \psi(t) \rangle|^2 = \lambda^{2L}, 
\end{align}
and thus
\begin{equation}
    f = -2 \log |\lambda|.
\end{equation}

%In Fig. \ref{fig:Potts-fE}, we show the fidelity and energy density as a function of the speed $\varv$, for multiple bond dimensions $\chi$, for the 3-state Potts model.

%\begin{figure}
%    \centering
%    \includegraphics[width=0.5\linewidth]{Potts-fE.pdf}
%    \caption{The fidelity, and excitation energy densities after a Kibble Zurek sweep performed at speed $\varv$ for the 3-state Potts model. We show the results for different maximum bond dimensions $\chi$. We show a black line illustrating the scaling prediction for $\chi=\infty$.}
%    \label{fig:Potts-fE}
%\end{figure}

\section{Best fit to extract $\kappa$}
\label{app:fit}

In the main text, the scaling function collapse depended on the length scale introduced by working at finite bond dimension. In previous work, it was demonstrated that the length scale is takes the form of $\xi_{\chi} \sim \chi^{\kappa}$ with $\kappa$ given by Eq. \eqref{eq:kappa}. Here, we find $\kappa$ by the best fit for the scaling function collapse for both the fidelity and excitation energy densities. Generically, the scaling ansatz for a Kibble-Zurek sweep at finite $\chi$ is given by Eq. \eqref{eq:n-chi}. If one fixes $\chi$ and $\varv$, then
\begin{equation}
    f_{\mathcal{O}}(\xi_{KZ} / \xi_{\chi}) = \mathcal{O}(\varv, \chi) / \mathcal{O}(\varv, \chi=\infty)
\end{equation}
For simplicity, define the right hand side as $y_{\chi}$, and the argument $x_{\kappa}:= \xi_{KZ} / \xi_{\chi}$. Then, for each $\chi$, our numerical data gives an interval $I_{\chi}$ for calculated values of $x$. We then interpolate the values for $y_{\chi} (x_{\kappa})$ for $x_{\kappa}\in I_{\chi}$. Call the interpolated function $\tilde y_{\chi}$. To find $\kappa$, we define a cost function
\begin{equation}\label{eq:cost}
    C(\kappa) = \sum_{\chi_i > \chi_j} \lVert\tilde{y}_{\chi_i}(x_\kappa) - \tilde{y}_{\chi_j}(x_{\kappa})\rVert^2,\quad x_{\kappa}\in I_{\chi_i}\cap I_{\chi_j}
\end{equation}
Conceptually, the cost function is given by the sum of the differences between all pairs of $\tilde{y}_{\chi_i},\tilde{y}_{\chi_j}$, along the interval where $x_\kappa$ is defined for both. We then find $\kappa$ by minimizing $C(\kappa)$. We show in Fig. \ref{fig:scaling-fit} the results of the best fit scaling function, as well as the extracted $\kappa$ values, for the Ising and 3-state Potts models.  Note that for both models the difference between best-fit and theoretical values is small, and comparable to the differences between best-fit $\kappa$ values taken from fidelity and energy results, which (unless KZ scaling is incorrect) should ultimately be the same in the $\chi \rightarrow \infty$ limit.  For the Ising case, the cost-function measure of error is essentially the same between the best fit and the theoretical value.

\begin{figure}
    \centering
    \includegraphics[width=0.49\linewidth]{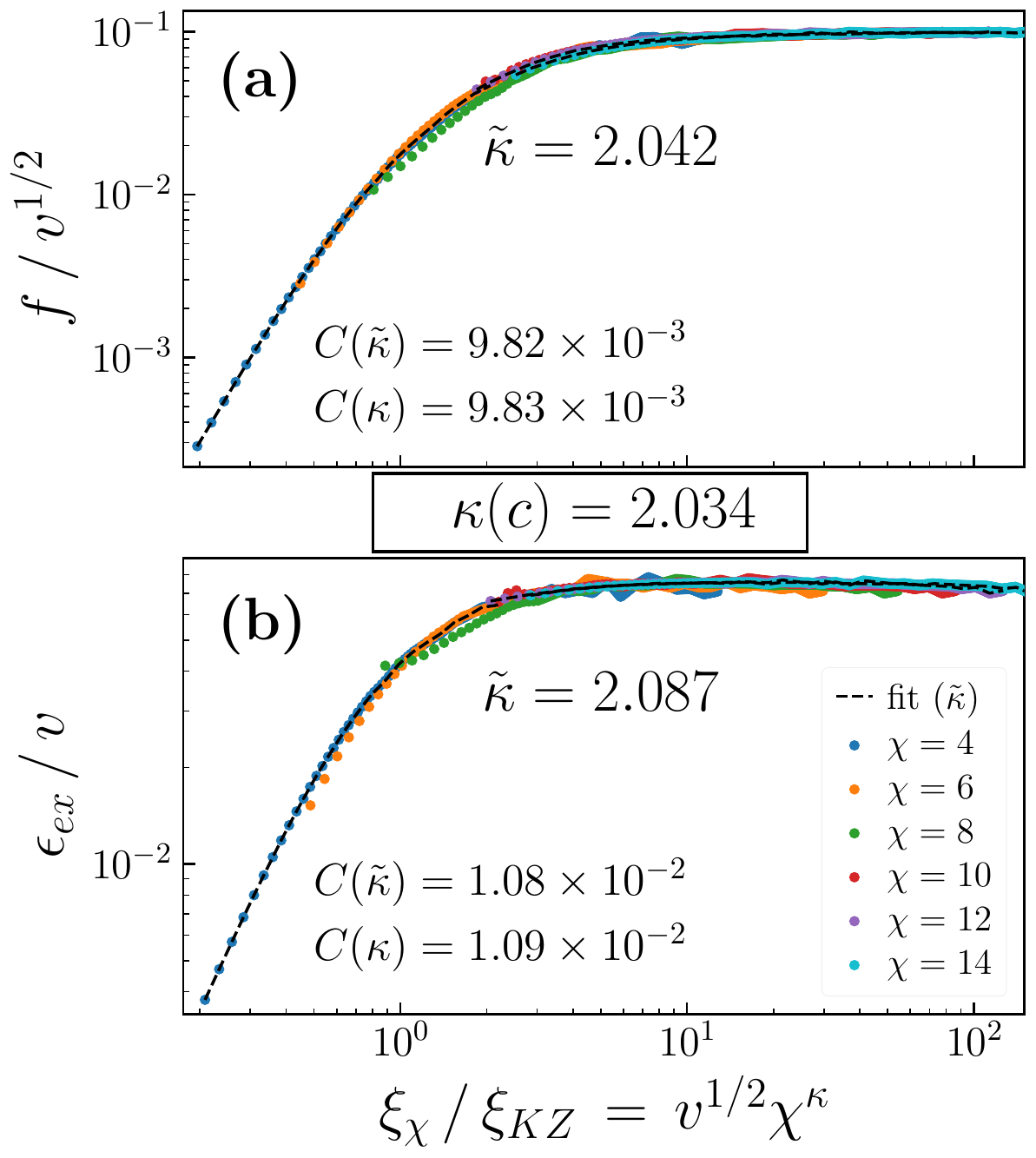}
    \includegraphics[width=0.49\linewidth]{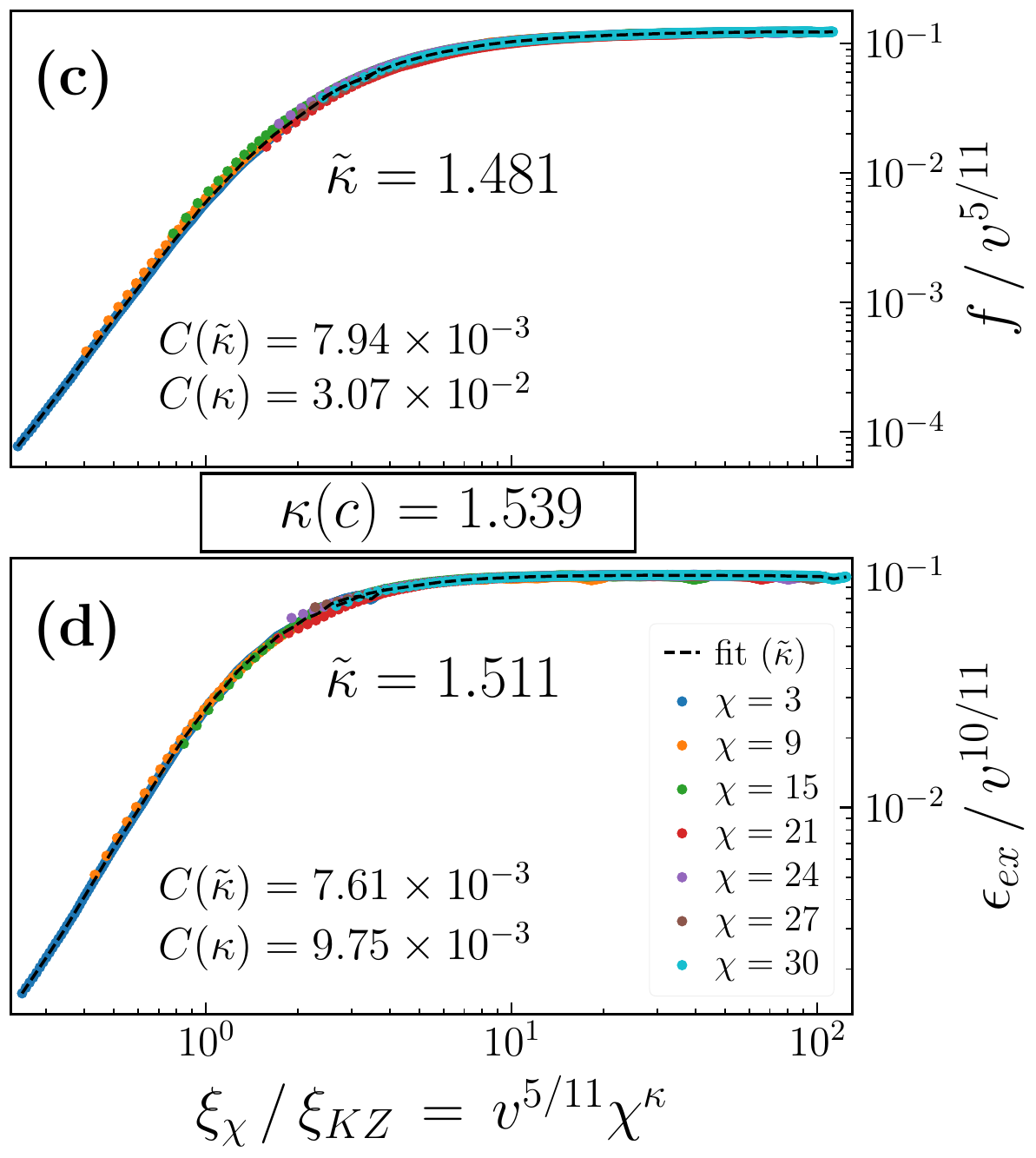}
    \caption{The best fit scaling function collapse for the fidelity and excitation energy densities, for the TFIM in (a) and (b), and the 3-state Potts model in (c) and (d). The value $\tilde{\kappa}$ is shown on the plots, and the scaling functions are drawn using that value of $\kappa$. The theoretical values are shown between the subplots as a reference. The values for the cost function of Eq. \eqref{eq:cost} are shown for the theoretical and best-fit values of $\kappa$. }
    \label{fig:scaling-fit}
\end{figure}

\section{Extrapolation to $\chi=\infty$}
\label{app:extrapolation}

One major benefit of the realization that observables satisfy the scaling relation in Eq. \eqref{eq:n-chi} is that it provides a means to extrapolate systematically to $\chi=\infty$ from finite-$\chi$ data. To see this, we can write
\begin{equation}\label{eq:extrapolate}
    \mathcal{O}(\varv, \chi=\infty) = 
    \mathcal{O}(\varv, \chi)
    \,/\,
    f_{\mathcal{O}}(\xi_{KZ} / \xi_{\chi}).
\end{equation}
The quantity $\mathcal{O}(\varv,\chi)$ is what is computed in the simulations. As for $f_{\mathcal{O}}$, this is computed by finding the function that the data collapses onto for different values of $\chi$. Where the collapse is not exact, we opt for the average over the different $\chi$ values. Then, the ratio of these two values yields the $\chi=\infty$ result, computed for each value of $\chi$. The resulting fidelity and excitation energy densities are shown in Fig. \ref{fig:extrapolation}, for both the TFIM, and the 3-state Potts model.

\begin{figure}
    \centering
    \includegraphics[width=0.49\linewidth]{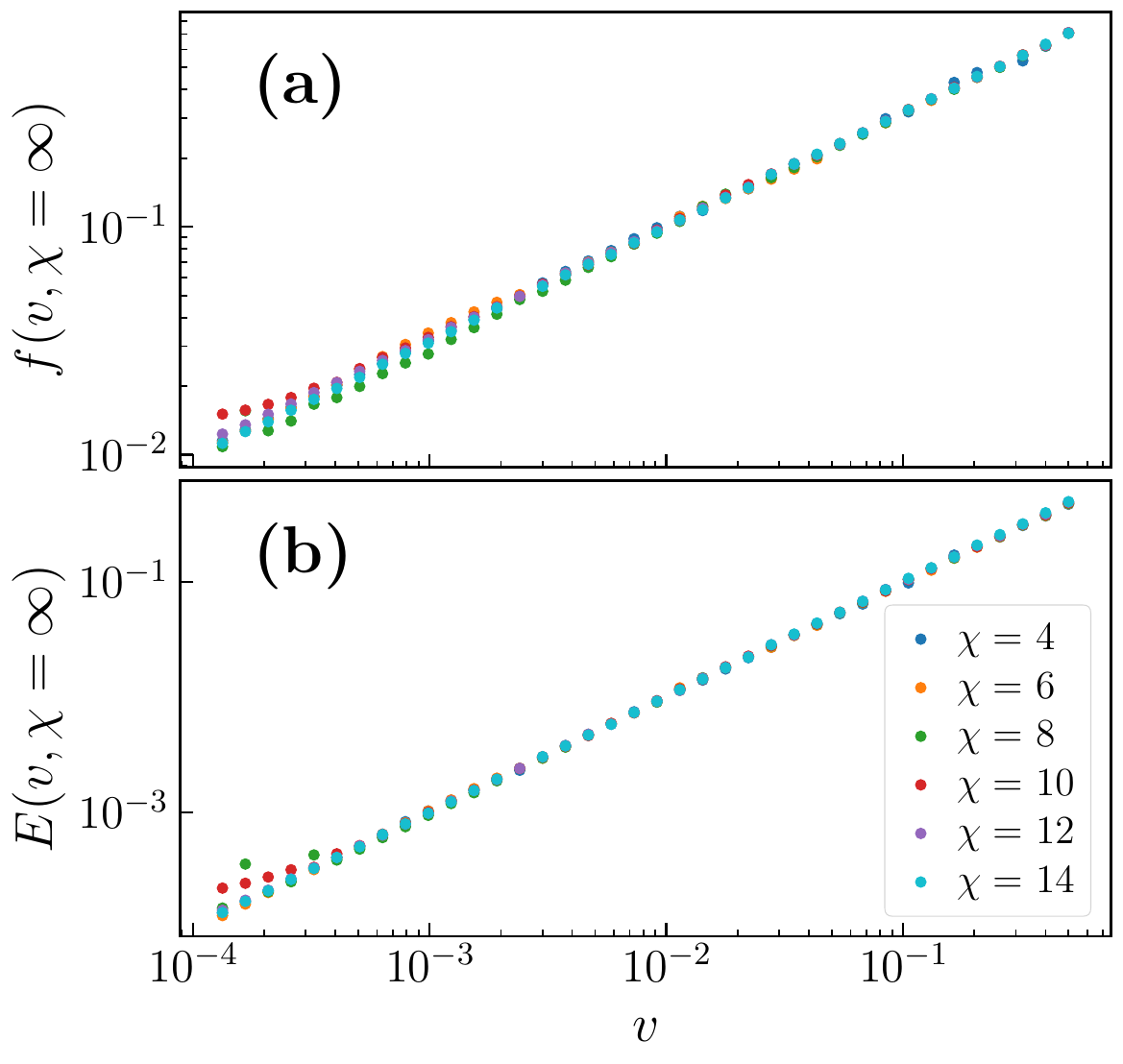}
    \includegraphics[width=0.49\linewidth]{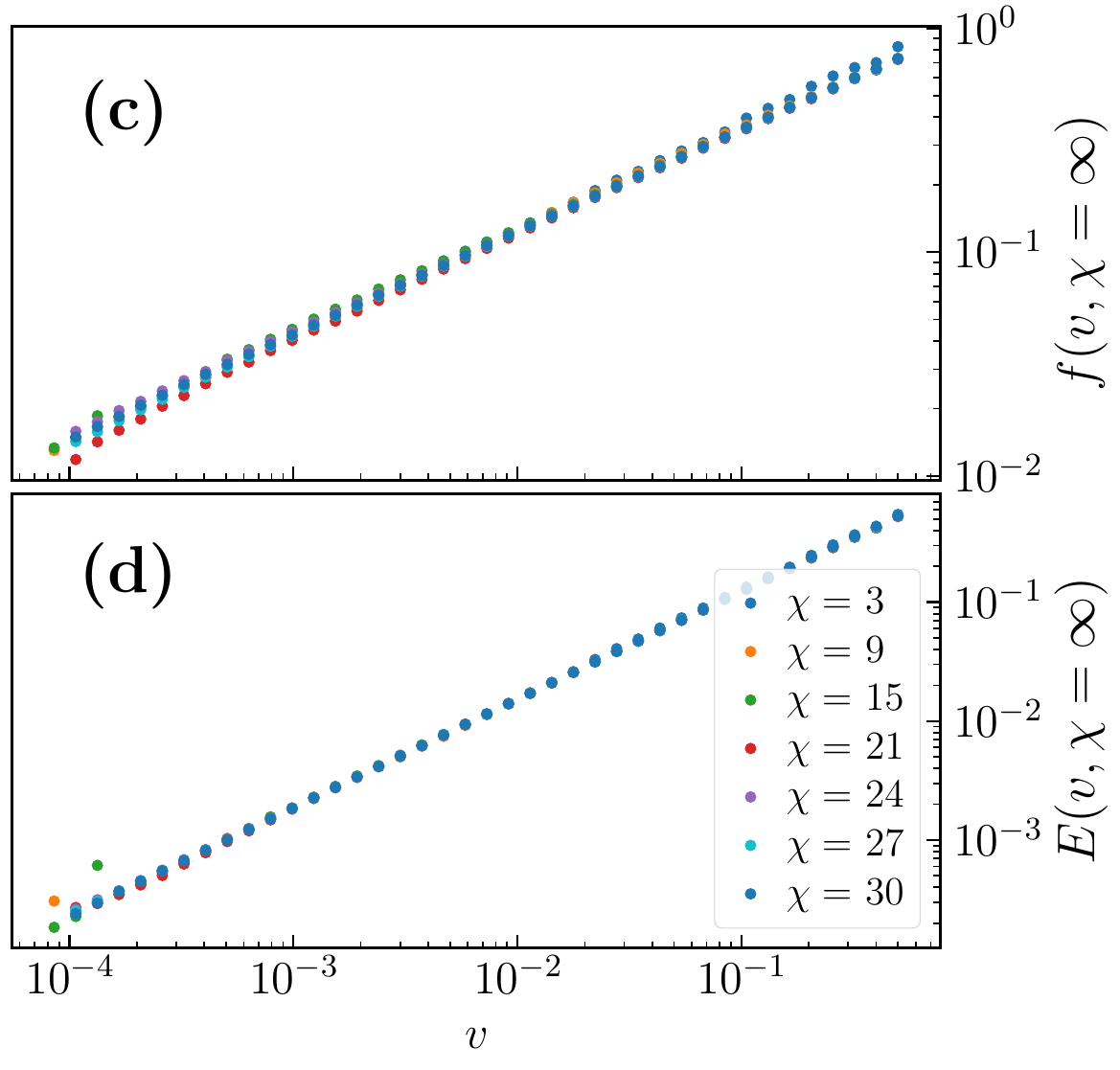}
    \caption{The extrapolated $\chi=\infty$ result for the fidelity and excitation energy densities, for the TFIM in (a) and (b), and the 3-state Potts model in (c) and (d). The extrapolation is computed using Eq. \eqref{eq:extrapolate}.}
    \label{fig:extrapolation}
\end{figure}

\end{widetext}

\bibliography{references}

\begin{thebibliography}{75}%
\makeatletter
\providecommand \@ifxundefined [1]{%
 \@ifx{#1\undefined}
}%
\providecommand \@ifnum [1]{%
 \ifnum #1\expandafter \@firstoftwo
 \else \expandafter \@secondoftwo
 \fi
}%
\providecommand \@ifx [1]{%
 \ifx #1\expandafter \@firstoftwo
 \else \expandafter \@secondoftwo
 \fi
}%
\providecommand \natexlab [1]{#1}%
\providecommand \enquote  [1]{``#1''}%
\providecommand \bibnamefont  [1]{#1}%
\providecommand \bibfnamefont [1]{#1}%
\providecommand \citenamefont [1]{#1}%
\providecommand \href@noop [0]{\@secondoftwo}%
\providecommand \href [0]{\begingroup \@sanitize@url \@href}%
\providecommand \@href[1]{\@@startlink{#1}\@@href}%
\providecommand \@@href[1]{\endgroup#1\@@endlink}%
\providecommand \@sanitize@url [0]{\catcode `\\12\catcode `\$12\catcode
  `\&12\catcode `\#12\catcode `\^12\catcode `\_12\catcode `\%12\relax}%
\providecommand \@@startlink[1]{}%
\providecommand \@@endlink[0]{}%
\providecommand \url  [0]{\begingroup\@sanitize@url \@url }%
\providecommand \@url [1]{\endgroup\@href {#1}{\urlprefix }}%
\providecommand \urlprefix  [0]{URL }%
\providecommand \Eprint [0]{\href }%
\providecommand \doibase [0]{https://doi.org/}%
\providecommand \selectlanguage [0]{\@gobble}%
\providecommand \bibinfo  [0]{\@secondoftwo}%
\providecommand \bibfield  [0]{\@secondoftwo}%
\providecommand \translation [1]{[#1]}%
\providecommand \BibitemOpen [0]{}%
\providecommand \bibitemStop [0]{}%
\providecommand \bibitemNoStop [0]{.\EOS\space}%
\providecommand \EOS [0]{\spacefactor3000\relax}%
\providecommand \BibitemShut  [1]{\csname bibitem#1\endcsname}%
\let\auto@bib@innerbib\@empty
%</preamble>
\bibitem [{\citenamefont {Calabrese}\ and\ \citenamefont
  {Cardy}(2005)}]{Calabrese_2005}%
  \BibitemOpen
  \bibfield  {author} {\bibinfo {author} {\bibfnamefont {P.}~\bibnamefont
  {Calabrese}}\ and\ \bibinfo {author} {\bibfnamefont {J.}~\bibnamefont
  {Cardy}},\ }\bibfield  {title} {\bibinfo {title} {Evolution of entanglement
  entropy in one-dimensional systems},\ }\href
  {https://doi.org/10.1088/1742-5468/2005/04/P04010} {\bibfield  {journal}
  {\bibinfo  {journal} {Journal of Statistical Mechanics: Theory and
  Experiment}\ }\textbf {\bibinfo {volume} {2005}},\ \bibinfo {pages} {P04010}
  (\bibinfo {year} {2005})}\BibitemShut {NoStop}%
\bibitem [{\citenamefont {Chiara}\ \emph {et~al.}(2006)\citenamefont {Chiara},
  \citenamefont {Montangero}, \citenamefont {Calabrese},\ and\ \citenamefont
  {Fazio}}]{De_Chiara_2006}%
  \BibitemOpen
  \bibfield  {author} {\bibinfo {author} {\bibfnamefont {G.~D.}\ \bibnamefont
  {Chiara}}, \bibinfo {author} {\bibfnamefont {S.}~\bibnamefont {Montangero}},
  \bibinfo {author} {\bibfnamefont {P.}~\bibnamefont {Calabrese}},\ and\
  \bibinfo {author} {\bibfnamefont {R.}~\bibnamefont {Fazio}},\ }\bibfield
  {title} {\bibinfo {title} {Entanglement entropy dynamics of heisenberg
  chains},\ }\href {https://doi.org/10.1088/1742-5468/2006/03/P03001}
  {\bibfield  {journal} {\bibinfo  {journal} {Journal of Statistical Mechanics:
  Theory and Experiment}\ }\textbf {\bibinfo {volume} {2006}},\ \bibinfo
  {pages} {P03001} (\bibinfo {year} {2006})}\BibitemShut {NoStop}%
\bibitem [{\citenamefont {Kim}\ and\ \citenamefont
  {Huse}(2013)}]{PhysRevLett.111.127205}%
  \BibitemOpen
  \bibfield  {author} {\bibinfo {author} {\bibfnamefont {H.}~\bibnamefont
  {Kim}}\ and\ \bibinfo {author} {\bibfnamefont {D.~A.}\ \bibnamefont {Huse}},\
  }\bibfield  {title} {\bibinfo {title} {Ballistic spreading of entanglement in
  a diffusive nonintegrable system},\ }\href
  {https://doi.org/10.1103/PhysRevLett.111.127205} {\bibfield  {journal}
  {\bibinfo  {journal} {Phys. Rev. Lett.}\ }\textbf {\bibinfo {volume} {111}},\
  \bibinfo {pages} {127205} (\bibinfo {year} {2013})}\BibitemShut {NoStop}%
\bibitem [{\citenamefont {Ho}\ and\ \citenamefont
  {Abanin}(2017)}]{PhysRevB.95.094302}%
  \BibitemOpen
  \bibfield  {author} {\bibinfo {author} {\bibfnamefont {W.~W.}\ \bibnamefont
  {Ho}}\ and\ \bibinfo {author} {\bibfnamefont {D.~A.}\ \bibnamefont
  {Abanin}},\ }\bibfield  {title} {\bibinfo {title} {Entanglement dynamics in
  quantum many-body systems},\ }\href
  {https://doi.org/10.1103/PhysRevB.95.094302} {\bibfield  {journal} {\bibinfo
  {journal} {Phys. Rev. B}\ }\textbf {\bibinfo {volume} {95}},\ \bibinfo
  {pages} {094302} (\bibinfo {year} {2017})}\BibitemShut {NoStop}%
\bibitem [{\citenamefont {Rigol}\ \emph {et~al.}(2008)\citenamefont {Rigol},
  \citenamefont {Dunjko},\ and\ \citenamefont {Olshanii}}]{Rigol_2008}%
  \BibitemOpen
  \bibfield  {author} {\bibinfo {author} {\bibfnamefont {M.}~\bibnamefont
  {Rigol}}, \bibinfo {author} {\bibfnamefont {V.}~\bibnamefont {Dunjko}},\ and\
  \bibinfo {author} {\bibfnamefont {M.}~\bibnamefont {Olshanii}},\ }\bibfield
  {title} {\bibinfo {title} {Thermalization and its mechanism for generic
  isolated quantum systems},\ }\href {https://doi.org/10.1038/nature06838}
  {\bibfield  {journal} {\bibinfo  {journal} {Nature}\ }\textbf {\bibinfo
  {volume} {452}},\ \bibinfo {pages} {854} (\bibinfo {year}
  {2008})}\BibitemShut {NoStop}%
\bibitem [{\citenamefont {D'Alessio}\ \emph {et~al.}(2016)\citenamefont
  {D'Alessio}, \citenamefont {Kafri}, \citenamefont {Polkovnikov},\ and\
  \citenamefont {Rigol}}]{DAlessio_2016}%
  \BibitemOpen
  \bibfield  {author} {\bibinfo {author} {\bibfnamefont {L.}~\bibnamefont
  {D'Alessio}}, \bibinfo {author} {\bibfnamefont {Y.}~\bibnamefont {Kafri}},
  \bibinfo {author} {\bibfnamefont {A.}~\bibnamefont {Polkovnikov}},\ and\
  \bibinfo {author} {\bibfnamefont {M.}~\bibnamefont {Rigol}},\ }\bibfield
  {title} {\bibinfo {title} {From quantum chaos and eigenstate thermalization
  to statistical mechanics and thermodynamics},\ }\href
  {https://doi.org/10.1080/00018732.2016.1198134} {\bibfield  {journal}
  {\bibinfo  {journal} {Advances in Physics}\ }\textbf {\bibinfo {volume}
  {65}},\ \bibinfo {pages} {239} (\bibinfo {year} {2016})},\ \Eprint
  {https://arxiv.org/abs/https://doi.org/10.1080/00018732.2016.1198134}
  {https://doi.org/10.1080/00018732.2016.1198134} \BibitemShut {NoStop}%
\bibitem [{\citenamefont {Ebadi}\ \emph {et~al.}(2021)\citenamefont {Ebadi},
  \citenamefont {Wang}, \citenamefont {Levine}, \citenamefont {Keesling},
  \citenamefont {Semeghini}, \citenamefont {Omran}, \citenamefont {Bluvstein},
  \citenamefont {Samajdar}, \citenamefont {Pichler}, \citenamefont {Ho},
  \citenamefont {Choi}, \citenamefont {Sachdev}, \citenamefont {Greiner},
  \citenamefont {Vuleti{\'c}},\ and\ \citenamefont {Lukin}}]{ebadi2021}%
  \BibitemOpen
  \bibfield  {author} {\bibinfo {author} {\bibfnamefont {S.}~\bibnamefont
  {Ebadi}}, \bibinfo {author} {\bibfnamefont {T.~T.}\ \bibnamefont {Wang}},
  \bibinfo {author} {\bibfnamefont {H.}~\bibnamefont {Levine}}, \bibinfo
  {author} {\bibfnamefont {A.}~\bibnamefont {Keesling}}, \bibinfo {author}
  {\bibfnamefont {G.}~\bibnamefont {Semeghini}}, \bibinfo {author}
  {\bibfnamefont {A.}~\bibnamefont {Omran}}, \bibinfo {author} {\bibfnamefont
  {D.}~\bibnamefont {Bluvstein}}, \bibinfo {author} {\bibfnamefont
  {R.}~\bibnamefont {Samajdar}}, \bibinfo {author} {\bibfnamefont
  {H.}~\bibnamefont {Pichler}}, \bibinfo {author} {\bibfnamefont {W.~W.}\
  \bibnamefont {Ho}}, \bibinfo {author} {\bibfnamefont {S.}~\bibnamefont
  {Choi}}, \bibinfo {author} {\bibfnamefont {S.}~\bibnamefont {Sachdev}},
  \bibinfo {author} {\bibfnamefont {M.}~\bibnamefont {Greiner}}, \bibinfo
  {author} {\bibfnamefont {V.}~\bibnamefont {Vuleti{\'c}}},\ and\ \bibinfo
  {author} {\bibfnamefont {M.~D.}\ \bibnamefont {Lukin}},\ }\bibfield  {title}
  {\bibinfo {title} {Quantum phases of matter on a 256-atom programmable
  quantum simulator},\ }\href {https://doi.org/10.1038/s41586-021-03582-4}
  {\bibfield  {journal} {\bibinfo  {journal} {Nature}\ }\textbf {\bibinfo
  {volume} {595}},\ \bibinfo {pages} {227} (\bibinfo {year}
  {2021})}\BibitemShut {NoStop}%
\bibitem [{car(1988)}]{cardyfss}%
  \BibitemOpen
  \bibfield  {title} {\bibinfo {title} {1 - introduction to theory of
  finite-size scaling},\ }in\ \href
  {https://doi.org/https://doi.org/10.1016/B978-0-444-87109-1.50006-6} {\emph
  {\bibinfo {booktitle} {Finite-Size Scaling}}},\ \bibinfo {series} {Current
  Physics–Sources and Comments}, Vol.~\bibinfo {volume} {2},\ \bibinfo
  {editor} {edited by\ \bibinfo {editor} {\bibfnamefont {J.~L.}\ \bibnamefont
  {CARDY}}}\ (\bibinfo  {publisher} {Elsevier},\ \bibinfo {year} {1988})\ pp.\
  \bibinfo {pages} {1--7}\BibitemShut {NoStop}%
\bibitem [{\citenamefont {Bernien}\ \emph {et~al.}(2017)\citenamefont
  {Bernien}, \citenamefont {Schwartz}, \citenamefont {Keesling}, \citenamefont
  {Levine}, \citenamefont {Omran}, \citenamefont {Pichler}, \citenamefont
  {Choi}, \citenamefont {Zibrov}, \citenamefont {Endres}, \citenamefont
  {Greiner}, \citenamefont {Vuleti{\'c}},\ and\ \citenamefont
  {Lukin}}]{Bernien_2017}%
  \BibitemOpen
  \bibfield  {author} {\bibinfo {author} {\bibfnamefont {H.}~\bibnamefont
  {Bernien}}, \bibinfo {author} {\bibfnamefont {S.}~\bibnamefont {Schwartz}},
  \bibinfo {author} {\bibfnamefont {A.}~\bibnamefont {Keesling}}, \bibinfo
  {author} {\bibfnamefont {H.}~\bibnamefont {Levine}}, \bibinfo {author}
  {\bibfnamefont {A.}~\bibnamefont {Omran}}, \bibinfo {author} {\bibfnamefont
  {H.}~\bibnamefont {Pichler}}, \bibinfo {author} {\bibfnamefont
  {S.}~\bibnamefont {Choi}}, \bibinfo {author} {\bibfnamefont {A.~S.}\
  \bibnamefont {Zibrov}}, \bibinfo {author} {\bibfnamefont {M.}~\bibnamefont
  {Endres}}, \bibinfo {author} {\bibfnamefont {M.}~\bibnamefont {Greiner}},
  \bibinfo {author} {\bibfnamefont {V.}~\bibnamefont {Vuleti{\'c}}},\ and\
  \bibinfo {author} {\bibfnamefont {M.~D.}\ \bibnamefont {Lukin}},\ }\bibfield
  {title} {\bibinfo {title} {Probing many-body dynamics on a 51-atom quantum
  simulator},\ }\href {https://doi.org/10.1038/nature24622} {\bibfield
  {journal} {\bibinfo  {journal} {Nature}\ }\textbf {\bibinfo {volume} {551}},\
  \bibinfo {pages} {579} (\bibinfo {year} {2017})}\BibitemShut {NoStop}%
\bibitem [{\citenamefont {Zhang}\ \emph {et~al.}(2017)\citenamefont {Zhang},
  \citenamefont {Pagano}, \citenamefont {Hess}, \citenamefont {Kyprianidis},
  \citenamefont {Becker}, \citenamefont {Kaplan}, \citenamefont {Gorshkov},
  \citenamefont {Gong},\ and\ \citenamefont {Monroe}}]{Zhang_2017}%
  \BibitemOpen
  \bibfield  {author} {\bibinfo {author} {\bibfnamefont {J.}~\bibnamefont
  {Zhang}}, \bibinfo {author} {\bibfnamefont {G.}~\bibnamefont {Pagano}},
  \bibinfo {author} {\bibfnamefont {P.~W.}\ \bibnamefont {Hess}}, \bibinfo
  {author} {\bibfnamefont {A.}~\bibnamefont {Kyprianidis}}, \bibinfo {author}
  {\bibfnamefont {P.}~\bibnamefont {Becker}}, \bibinfo {author} {\bibfnamefont
  {H.}~\bibnamefont {Kaplan}}, \bibinfo {author} {\bibfnamefont {A.~V.}\
  \bibnamefont {Gorshkov}}, \bibinfo {author} {\bibfnamefont {Z.~X.}\
  \bibnamefont {Gong}},\ and\ \bibinfo {author} {\bibfnamefont
  {C.}~\bibnamefont {Monroe}},\ }\bibfield  {title} {\bibinfo {title}
  {Observation of a many-body dynamical phase transition with a 53-qubit
  quantum simulator},\ }\href {https://doi.org/10.1038/nature24654} {\bibfield
  {journal} {\bibinfo  {journal} {Nature}\ }\textbf {\bibinfo {volume} {551}},\
  \bibinfo {pages} {601} (\bibinfo {year} {2017})}\BibitemShut {NoStop}%
\bibitem [{\citenamefont {Dupont}\ and\ \citenamefont
  {Moore}(2022)}]{PhysRevB.106.L041109}%
  \BibitemOpen
  \bibfield  {author} {\bibinfo {author} {\bibfnamefont {M.}~\bibnamefont
  {Dupont}}\ and\ \bibinfo {author} {\bibfnamefont {J.~E.}\ \bibnamefont
  {Moore}},\ }\bibfield  {title} {\bibinfo {title} {Quantum criticality using a
  superconducting quantum processor},\ }\href
  {https://doi.org/10.1103/PhysRevB.106.L041109} {\bibfield  {journal}
  {\bibinfo  {journal} {Phys. Rev. B}\ }\textbf {\bibinfo {volume} {106}},\
  \bibinfo {pages} {L041109} (\bibinfo {year} {2022})}\BibitemShut {NoStop}%
\bibitem [{\citenamefont {Rams}\ \emph {et~al.}(2019)\citenamefont {Rams},
  \citenamefont {Dziarmaga},\ and\ \citenamefont {Zurek}}]{rams2019symmetry}%
  \BibitemOpen
  \bibfield  {author} {\bibinfo {author} {\bibfnamefont {M.~M.}\ \bibnamefont
  {Rams}}, \bibinfo {author} {\bibfnamefont {J.}~\bibnamefont {Dziarmaga}},\
  and\ \bibinfo {author} {\bibfnamefont {W.~H.}\ \bibnamefont {Zurek}},\
  }\bibfield  {title} {\bibinfo {title} {Symmetry breaking bias and the
  dynamics of a quantum phase transition},\ }\href
  {https://doi.org/10.1103/PhysRevLett.123.130603} {\bibfield  {journal}
  {\bibinfo  {journal} {Phys. Rev. Lett.}\ }\textbf {\bibinfo {volume} {123}},\
  \bibinfo {pages} {130603} (\bibinfo {year} {2019})}\BibitemShut {NoStop}%
\bibitem [{\citenamefont {Hódsági}\ and\ \citenamefont
  {Kormos}(2020)}]{10.21468/SciPostPhys.9.4.055}%
  \BibitemOpen
  \bibfield  {author} {\bibinfo {author} {\bibfnamefont {K.}~\bibnamefont
  {Hódsági}}\ and\ \bibinfo {author} {\bibfnamefont {M.}~\bibnamefont
  {Kormos}},\ }\bibfield  {title} {\bibinfo {title} {{Kibble–Zurek mechanism
  in the Ising Field Theory}},\ }\href
  {https://doi.org/10.21468/SciPostPhys.9.4.055} {\bibfield  {journal}
  {\bibinfo  {journal} {SciPost Phys.}\ }\textbf {\bibinfo {volume} {9}},\
  \bibinfo {pages} {055} (\bibinfo {year} {2020})}\BibitemShut {NoStop}%
\bibitem [{\citenamefont {White}(1992)}]{White1992}%
  \BibitemOpen
  \bibfield  {author} {\bibinfo {author} {\bibfnamefont {S.~R.}\ \bibnamefont
  {White}},\ }\bibfield  {title} {\bibinfo {title} {Density matrix formulation
  for quantum renormalization groups},\ }\href
  {https://doi.org/10.1103/PhysRevLett.69.2863} {\bibfield  {journal} {\bibinfo
   {journal} {Phys. Rev. Lett.}\ }\textbf {\bibinfo {volume} {69}},\ \bibinfo
  {pages} {2863} (\bibinfo {year} {1992})}\BibitemShut {NoStop}%
\bibitem [{\citenamefont {Schollwöck}(2011)}]{Schollwock2011}%
  \BibitemOpen
  \bibfield  {author} {\bibinfo {author} {\bibfnamefont {U.}~\bibnamefont
  {Schollwöck}},\ }\bibfield  {title} {\bibinfo {title} {The density-matrix
  renormalization group in the age of matrix product states},\ }\href
  {https://doi.org/https://doi.org/10.1016/j.aop.2010.09.012} {\bibfield
  {journal} {\bibinfo  {journal} {Annals of Physics}\ }\textbf {\bibinfo
  {volume} {326}},\ \bibinfo {pages} {96} (\bibinfo {year} {2011})},\ \bibinfo
  {note} {january 2011 Special Issue}\BibitemShut {NoStop}%
\bibitem [{\citenamefont {Hastings}(2007)}]{Hastings_2007}%
  \BibitemOpen
  \bibfield  {author} {\bibinfo {author} {\bibfnamefont {M.~B.}\ \bibnamefont
  {Hastings}},\ }\bibfield  {title} {\bibinfo {title} {An area law for
  one-dimensional quantum systems},\ }\href
  {https://doi.org/10.1088/1742-5468/2007/08/p08024} {\bibfield  {journal}
  {\bibinfo  {journal} {Journal of Statistical Mechanics: Theory and
  Experiment}\ }\textbf {\bibinfo {volume} {2007}},\ \bibinfo {pages} {P08024}
  (\bibinfo {year} {2007})}\BibitemShut {NoStop}%
\bibitem [{\citenamefont {Verstraete}\ and\ \citenamefont
  {Cirac}(2006)}]{PhysRevB.73.094423}%
  \BibitemOpen
  \bibfield  {author} {\bibinfo {author} {\bibfnamefont {F.}~\bibnamefont
  {Verstraete}}\ and\ \bibinfo {author} {\bibfnamefont {J.~I.}\ \bibnamefont
  {Cirac}},\ }\bibfield  {title} {\bibinfo {title} {Matrix product states
  represent ground states faithfully},\ }\href
  {https://doi.org/10.1103/PhysRevB.73.094423} {\bibfield  {journal} {\bibinfo
  {journal} {Phys. Rev. B}\ }\textbf {\bibinfo {volume} {73}},\ \bibinfo
  {pages} {094423} (\bibinfo {year} {2006})}\BibitemShut {NoStop}%
\bibitem [{\citenamefont {Osborne}(2007)}]{PhysRevA.75.042306}%
  \BibitemOpen
  \bibfield  {author} {\bibinfo {author} {\bibfnamefont {T.~J.}\ \bibnamefont
  {Osborne}},\ }\bibfield  {title} {\bibinfo {title} {Ground state of a class
  of noncritical one-dimensional quantum spin systems can be approximated
  efficiently},\ }\href {https://doi.org/10.1103/PhysRevA.75.042306} {\bibfield
   {journal} {\bibinfo  {journal} {Phys. Rev. A}\ }\textbf {\bibinfo {volume}
  {75}},\ \bibinfo {pages} {042306} (\bibinfo {year} {2007})}\BibitemShut
  {NoStop}%
\bibitem [{\citenamefont {Vidal}(2007{\natexlab{a}})}]{PhysRevLett.98.070201}%
  \BibitemOpen
  \bibfield  {author} {\bibinfo {author} {\bibfnamefont {G.}~\bibnamefont
  {Vidal}},\ }\bibfield  {title} {\bibinfo {title} {Classical simulation of
  infinite-size quantum lattice systems in one spatial dimension},\ }\href
  {https://doi.org/10.1103/PhysRevLett.98.070201} {\bibfield  {journal}
  {\bibinfo  {journal} {Phys. Rev. Lett.}\ }\textbf {\bibinfo {volume} {98}},\
  \bibinfo {pages} {070201} (\bibinfo {year} {2007}{\natexlab{a}})}\BibitemShut
  {NoStop}%
\bibitem [{\citenamefont {Vidal}(2007{\natexlab{b}})}]{PhysRevLett.99.220405}%
  \BibitemOpen
  \bibfield  {author} {\bibinfo {author} {\bibfnamefont {G.}~\bibnamefont
  {Vidal}},\ }\bibfield  {title} {\bibinfo {title} {Entanglement
  renormalization},\ }\href {https://doi.org/10.1103/PhysRevLett.99.220405}
  {\bibfield  {journal} {\bibinfo  {journal} {Phys. Rev. Lett.}\ }\textbf
  {\bibinfo {volume} {99}},\ \bibinfo {pages} {220405} (\bibinfo {year}
  {2007}{\natexlab{b}})}\BibitemShut {NoStop}%
\bibitem [{\citenamefont {Vidal}(2008)}]{PhysRevLett.101.110501}%
  \BibitemOpen
  \bibfield  {author} {\bibinfo {author} {\bibfnamefont {G.}~\bibnamefont
  {Vidal}},\ }\bibfield  {title} {\bibinfo {title} {Class of quantum many-body
  states that can be efficiently simulated},\ }\href
  {https://doi.org/10.1103/PhysRevLett.101.110501} {\bibfield  {journal}
  {\bibinfo  {journal} {Phys. Rev. Lett.}\ }\textbf {\bibinfo {volume} {101}},\
  \bibinfo {pages} {110501} (\bibinfo {year} {2008})}\BibitemShut {NoStop}%
\bibitem [{\citenamefont {Verstraete}\ and\ \citenamefont
  {Cirac}(2004)}]{Verstraete2004}%
  \BibitemOpen
  \bibfield  {author} {\bibinfo {author} {\bibfnamefont {F.}~\bibnamefont
  {Verstraete}}\ and\ \bibinfo {author} {\bibfnamefont {J.~I.}\ \bibnamefont
  {Cirac}},\ }\href {https://doi.org/10.48550/ARXIV.COND-MAT/0407066} {\bibinfo
  {title} {Renormalization algorithms for quantum-many body systems in two and
  higher dimensions}} (\bibinfo {year} {2004})\BibitemShut {NoStop}%
\bibitem [{\citenamefont {Orús}(2014)}]{ORUS2014117}%
  \BibitemOpen
  \bibfield  {author} {\bibinfo {author} {\bibfnamefont {R.}~\bibnamefont
  {Orús}},\ }\bibfield  {title} {\bibinfo {title} {A practical introduction to
  tensor networks: Matrix product states and projected entangled pair states},\
  }\href {https://doi.org/https://doi.org/10.1016/j.aop.2014.06.013} {\bibfield
   {journal} {\bibinfo  {journal} {Annals of Physics}\ }\textbf {\bibinfo
  {volume} {349}},\ \bibinfo {pages} {117} (\bibinfo {year}
  {2014})}\BibitemShut {NoStop}%
\bibitem [{\citenamefont {Cirac}\ \emph {et~al.}(2019)\citenamefont {Cirac},
  \citenamefont {Garre-Rubio},\ and\ \citenamefont
  {P{\'e}rez-Garc{\'\i}a}}]{Cirac2019}%
  \BibitemOpen
  \bibfield  {author} {\bibinfo {author} {\bibfnamefont {J.~I.}\ \bibnamefont
  {Cirac}}, \bibinfo {author} {\bibfnamefont {J.}~\bibnamefont {Garre-Rubio}},\
  and\ \bibinfo {author} {\bibfnamefont {D.}~\bibnamefont
  {P{\'e}rez-Garc{\'\i}a}},\ }\bibfield  {title} {\bibinfo {title}
  {Mathematical open problems in projected entangled pair states},\ }\href@noop
  {} {\bibfield  {journal} {\bibinfo  {journal} {Revista Matem{\'a}tica
  Complutense}\ }\textbf {\bibinfo {volume} {32}},\ \bibinfo {pages} {579}
  (\bibinfo {year} {2019})}\BibitemShut {NoStop}%
\bibitem [{\citenamefont {Cirac}\ \emph {et~al.}(2021)\citenamefont {Cirac},
  \citenamefont {P\'erez-Garc\'{\i}a}, \citenamefont {Schuch},\ and\
  \citenamefont {Verstraete}}]{Cirac2021}%
  \BibitemOpen
  \bibfield  {author} {\bibinfo {author} {\bibfnamefont {J.~I.}\ \bibnamefont
  {Cirac}}, \bibinfo {author} {\bibfnamefont {D.}~\bibnamefont
  {P\'erez-Garc\'{\i}a}}, \bibinfo {author} {\bibfnamefont {N.}~\bibnamefont
  {Schuch}},\ and\ \bibinfo {author} {\bibfnamefont {F.}~\bibnamefont
  {Verstraete}},\ }\bibfield  {title} {\bibinfo {title} {Matrix product states
  and projected entangled pair states: Concepts, symmetries, theorems},\ }\href
  {https://doi.org/10.1103/RevModPhys.93.045003} {\bibfield  {journal}
  {\bibinfo  {journal} {Rev. Mod. Phys.}\ }\textbf {\bibinfo {volume} {93}},\
  \bibinfo {pages} {045003} (\bibinfo {year} {2021})}\BibitemShut {NoStop}%
\bibitem [{\citenamefont {\"Ostlund}\ and\ \citenamefont
  {Rommer}(1995)}]{PhysRevLett.75.3537}%
  \BibitemOpen
  \bibfield  {author} {\bibinfo {author} {\bibfnamefont {S.}~\bibnamefont
  {\"Ostlund}}\ and\ \bibinfo {author} {\bibfnamefont {S.}~\bibnamefont
  {Rommer}},\ }\bibfield  {title} {\bibinfo {title} {Thermodynamic limit of
  density matrix renormalization},\ }\href
  {https://doi.org/10.1103/PhysRevLett.75.3537} {\bibfield  {journal} {\bibinfo
   {journal} {Phys. Rev. Lett.}\ }\textbf {\bibinfo {volume} {75}},\ \bibinfo
  {pages} {3537} (\bibinfo {year} {1995})}\BibitemShut {NoStop}%
\bibitem [{\citenamefont {Rommer}\ and\ \citenamefont
  {\"Ostlund}(1997)}]{PhysRevB.55.2164}%
  \BibitemOpen
  \bibfield  {author} {\bibinfo {author} {\bibfnamefont {S.}~\bibnamefont
  {Rommer}}\ and\ \bibinfo {author} {\bibfnamefont {S.}~\bibnamefont
  {\"Ostlund}},\ }\bibfield  {title} {\bibinfo {title} {Class of ansatz wave
  functions for one-dimensional spin systems and their relation to the density
  matrix renormalization group},\ }\href
  {https://doi.org/10.1103/PhysRevB.55.2164} {\bibfield  {journal} {\bibinfo
  {journal} {Phys. Rev. B}\ }\textbf {\bibinfo {volume} {55}},\ \bibinfo
  {pages} {2164} (\bibinfo {year} {1997})}\BibitemShut {NoStop}%
\bibitem [{\citenamefont {Haegeman}\ \emph {et~al.}(2012)\citenamefont
  {Haegeman}, \citenamefont {Pirvu}, \citenamefont {Weir}, \citenamefont
  {Cirac}, \citenamefont {Osborne}, \citenamefont {Verschelde},\ and\
  \citenamefont {Verstraete}}]{PhysRevB.85.100408}%
  \BibitemOpen
  \bibfield  {author} {\bibinfo {author} {\bibfnamefont {J.}~\bibnamefont
  {Haegeman}}, \bibinfo {author} {\bibfnamefont {B.}~\bibnamefont {Pirvu}},
  \bibinfo {author} {\bibfnamefont {D.~J.}\ \bibnamefont {Weir}}, \bibinfo
  {author} {\bibfnamefont {J.~I.}\ \bibnamefont {Cirac}}, \bibinfo {author}
  {\bibfnamefont {T.~J.}\ \bibnamefont {Osborne}}, \bibinfo {author}
  {\bibfnamefont {H.}~\bibnamefont {Verschelde}},\ and\ \bibinfo {author}
  {\bibfnamefont {F.}~\bibnamefont {Verstraete}},\ }\bibfield  {title}
  {\bibinfo {title} {Variational matrix product ansatz for dispersion
  relations},\ }\href {https://doi.org/10.1103/PhysRevB.85.100408} {\bibfield
  {journal} {\bibinfo  {journal} {Phys. Rev. B}\ }\textbf {\bibinfo {volume}
  {85}},\ \bibinfo {pages} {100408} (\bibinfo {year} {2012})}\BibitemShut
  {NoStop}%
\bibitem [{\citenamefont {Haegeman}\ \emph
  {et~al.}(2013{\natexlab{a}})\citenamefont {Haegeman}, \citenamefont
  {Michalakis}, \citenamefont {Nachtergaele}, \citenamefont {Osborne},
  \citenamefont {Schuch},\ and\ \citenamefont
  {Verstraete}}]{PhysRevLett.111.080401}%
  \BibitemOpen
  \bibfield  {author} {\bibinfo {author} {\bibfnamefont {J.}~\bibnamefont
  {Haegeman}}, \bibinfo {author} {\bibfnamefont {S.}~\bibnamefont
  {Michalakis}}, \bibinfo {author} {\bibfnamefont {B.}~\bibnamefont
  {Nachtergaele}}, \bibinfo {author} {\bibfnamefont {T.~J.}\ \bibnamefont
  {Osborne}}, \bibinfo {author} {\bibfnamefont {N.}~\bibnamefont {Schuch}},\
  and\ \bibinfo {author} {\bibfnamefont {F.}~\bibnamefont {Verstraete}},\
  }\bibfield  {title} {\bibinfo {title} {Elementary excitations in gapped
  quantum spin systems},\ }\href
  {https://doi.org/10.1103/PhysRevLett.111.080401} {\bibfield  {journal}
  {\bibinfo  {journal} {Phys. Rev. Lett.}\ }\textbf {\bibinfo {volume} {111}},\
  \bibinfo {pages} {080401} (\bibinfo {year} {2013}{\natexlab{a}})}\BibitemShut
  {NoStop}%
\bibitem [{\citenamefont {Zauner-Stauber}\ \emph {et~al.}(2018)\citenamefont
  {Zauner-Stauber}, \citenamefont {Vanderstraeten}, \citenamefont {Haegeman},
  \citenamefont {McCulloch},\ and\ \citenamefont
  {Verstraete}}]{PhysRevB.97.235155}%
  \BibitemOpen
  \bibfield  {author} {\bibinfo {author} {\bibfnamefont {V.}~\bibnamefont
  {Zauner-Stauber}}, \bibinfo {author} {\bibfnamefont {L.}~\bibnamefont
  {Vanderstraeten}}, \bibinfo {author} {\bibfnamefont {J.}~\bibnamefont
  {Haegeman}}, \bibinfo {author} {\bibfnamefont {I.~P.}\ \bibnamefont
  {McCulloch}},\ and\ \bibinfo {author} {\bibfnamefont {F.}~\bibnamefont
  {Verstraete}},\ }\bibfield  {title} {\bibinfo {title} {Topological nature of
  spinons and holons: Elementary excitations from matrix product states with
  conserved symmetries},\ }\href {https://doi.org/10.1103/PhysRevB.97.235155}
  {\bibfield  {journal} {\bibinfo  {journal} {Phys. Rev. B}\ }\textbf {\bibinfo
  {volume} {97}},\ \bibinfo {pages} {235155} (\bibinfo {year}
  {2018})}\BibitemShut {NoStop}%
\bibitem [{\citenamefont {Haegeman}\ \emph
  {et~al.}(2013{\natexlab{b}})\citenamefont {Haegeman}, \citenamefont
  {Osborne},\ and\ \citenamefont {Verstraete}}]{PhysRevB.88.075133}%
  \BibitemOpen
  \bibfield  {author} {\bibinfo {author} {\bibfnamefont {J.}~\bibnamefont
  {Haegeman}}, \bibinfo {author} {\bibfnamefont {T.~J.}\ \bibnamefont
  {Osborne}},\ and\ \bibinfo {author} {\bibfnamefont {F.}~\bibnamefont
  {Verstraete}},\ }\bibfield  {title} {\bibinfo {title} {Post-matrix product
  state methods: To tangent space and beyond},\ }\href
  {https://doi.org/10.1103/PhysRevB.88.075133} {\bibfield  {journal} {\bibinfo
  {journal} {Phys. Rev. B}\ }\textbf {\bibinfo {volume} {88}},\ \bibinfo
  {pages} {075133} (\bibinfo {year} {2013}{\natexlab{b}})}\BibitemShut
  {NoStop}%
\bibitem [{\citenamefont {Haegeman}\ \emph {et~al.}(2011)\citenamefont
  {Haegeman}, \citenamefont {Cirac}, \citenamefont {Osborne}, \citenamefont
  {Pi\ifmmode~\check{z}\else \v{z}\fi{}orn}, \citenamefont {Verschelde},\ and\
  \citenamefont {Verstraete}}]{PhysRevLett.107.070601}%
  \BibitemOpen
  \bibfield  {author} {\bibinfo {author} {\bibfnamefont {J.}~\bibnamefont
  {Haegeman}}, \bibinfo {author} {\bibfnamefont {J.~I.}\ \bibnamefont {Cirac}},
  \bibinfo {author} {\bibfnamefont {T.~J.}\ \bibnamefont {Osborne}}, \bibinfo
  {author} {\bibfnamefont {I.}~\bibnamefont {Pi\ifmmode~\check{z}\else
  \v{z}\fi{}orn}}, \bibinfo {author} {\bibfnamefont {H.}~\bibnamefont
  {Verschelde}},\ and\ \bibinfo {author} {\bibfnamefont {F.}~\bibnamefont
  {Verstraete}},\ }\bibfield  {title} {\bibinfo {title} {Time-dependent
  variational principle for quantum lattices},\ }\href
  {https://doi.org/10.1103/PhysRevLett.107.070601} {\bibfield  {journal}
  {\bibinfo  {journal} {Phys. Rev. Lett.}\ }\textbf {\bibinfo {volume} {107}},\
  \bibinfo {pages} {070601} (\bibinfo {year} {2011})}\BibitemShut {NoStop}%
\bibitem [{\citenamefont {Haegeman}\ \emph {et~al.}(2016)\citenamefont
  {Haegeman}, \citenamefont {Lubich}, \citenamefont {Oseledets}, \citenamefont
  {Vandereycken},\ and\ \citenamefont {Verstraete}}]{PhysRevB.94.165116}%
  \BibitemOpen
  \bibfield  {author} {\bibinfo {author} {\bibfnamefont {J.}~\bibnamefont
  {Haegeman}}, \bibinfo {author} {\bibfnamefont {C.}~\bibnamefont {Lubich}},
  \bibinfo {author} {\bibfnamefont {I.}~\bibnamefont {Oseledets}}, \bibinfo
  {author} {\bibfnamefont {B.}~\bibnamefont {Vandereycken}},\ and\ \bibinfo
  {author} {\bibfnamefont {F.}~\bibnamefont {Verstraete}},\ }\bibfield  {title}
  {\bibinfo {title} {Unifying time evolution and optimization with matrix
  product states},\ }\href {https://doi.org/10.1103/PhysRevB.94.165116}
  {\bibfield  {journal} {\bibinfo  {journal} {Phys. Rev. B}\ }\textbf {\bibinfo
  {volume} {94}},\ \bibinfo {pages} {165116} (\bibinfo {year}
  {2016})}\BibitemShut {NoStop}%
\bibitem [{\citenamefont {Vanderstraeten}\ \emph
  {et~al.}(2019{\natexlab{a}})\citenamefont {Vanderstraeten}, \citenamefont
  {Haegeman},\ and\ \citenamefont {Verstraete}}]{Vanderstraeten2019}%
  \BibitemOpen
  \bibfield  {author} {\bibinfo {author} {\bibfnamefont {L.}~\bibnamefont
  {Vanderstraeten}}, \bibinfo {author} {\bibfnamefont {J.}~\bibnamefont
  {Haegeman}},\ and\ \bibinfo {author} {\bibfnamefont {F.}~\bibnamefont
  {Verstraete}},\ }\bibfield  {title} {\bibinfo {title} {{Tangent-space methods
  for uniform matrix product states}},\ }\href
  {https://doi.org/10.21468/SciPostPhysLectNotes.7} {\bibfield  {journal}
  {\bibinfo  {journal} {SciPost Phys. Lect. Notes}\ ,\ \bibinfo {pages} {7}}
  (\bibinfo {year} {2019}{\natexlab{a}})}\BibitemShut {NoStop}%
\bibitem [{\citenamefont {Paeckel}\ \emph {et~al.}(2019)\citenamefont
  {Paeckel}, \citenamefont {Köhler}, \citenamefont {Swoboda}, \citenamefont
  {Manmana}, \citenamefont {Schollwöck},\ and\ \citenamefont
  {Hubig}}]{PAECKEL2019167998}%
  \BibitemOpen
  \bibfield  {author} {\bibinfo {author} {\bibfnamefont {S.}~\bibnamefont
  {Paeckel}}, \bibinfo {author} {\bibfnamefont {T.}~\bibnamefont {Köhler}},
  \bibinfo {author} {\bibfnamefont {A.}~\bibnamefont {Swoboda}}, \bibinfo
  {author} {\bibfnamefont {S.~R.}\ \bibnamefont {Manmana}}, \bibinfo {author}
  {\bibfnamefont {U.}~\bibnamefont {Schollwöck}},\ and\ \bibinfo {author}
  {\bibfnamefont {C.}~\bibnamefont {Hubig}},\ }\bibfield  {title} {\bibinfo
  {title} {Time-evolution methods for matrix-product states},\ }\href
  {https://doi.org/https://doi.org/10.1016/j.aop.2019.167998} {\bibfield
  {journal} {\bibinfo  {journal} {Annals of Physics}\ }\textbf {\bibinfo
  {volume} {411}},\ \bibinfo {pages} {167998} (\bibinfo {year}
  {2019})}\BibitemShut {NoStop}%
\bibitem [{\citenamefont {Zaletel}\ \emph {et~al.}(2015)\citenamefont
  {Zaletel}, \citenamefont {Mong}, \citenamefont {Karrasch}, \citenamefont
  {Moore},\ and\ \citenamefont {Pollmann}}]{PhysRevB.91.165112}%
  \BibitemOpen
  \bibfield  {author} {\bibinfo {author} {\bibfnamefont {M.~P.}\ \bibnamefont
  {Zaletel}}, \bibinfo {author} {\bibfnamefont {R.~S.~K.}\ \bibnamefont
  {Mong}}, \bibinfo {author} {\bibfnamefont {C.}~\bibnamefont {Karrasch}},
  \bibinfo {author} {\bibfnamefont {J.~E.}\ \bibnamefont {Moore}},\ and\
  \bibinfo {author} {\bibfnamefont {F.}~\bibnamefont {Pollmann}},\ }\bibfield
  {title} {\bibinfo {title} {Time-evolving a matrix product state with
  long-ranged interactions},\ }\href
  {https://doi.org/10.1103/PhysRevB.91.165112} {\bibfield  {journal} {\bibinfo
  {journal} {Phys. Rev. B}\ }\textbf {\bibinfo {volume} {91}},\ \bibinfo
  {pages} {165112} (\bibinfo {year} {2015})}\BibitemShut {NoStop}%
\bibitem [{\citenamefont {Daley}\ \emph {et~al.}(2004)\citenamefont {Daley},
  \citenamefont {Kollath}, \citenamefont {Schollwöck},\ and\ \citenamefont
  {Vidal}}]{Daley_2004}%
  \BibitemOpen
  \bibfield  {author} {\bibinfo {author} {\bibfnamefont {A.~J.}\ \bibnamefont
  {Daley}}, \bibinfo {author} {\bibfnamefont {C.}~\bibnamefont {Kollath}},
  \bibinfo {author} {\bibfnamefont {U.}~\bibnamefont {Schollwöck}},\ and\
  \bibinfo {author} {\bibfnamefont {G.}~\bibnamefont {Vidal}},\ }\bibfield
  {title} {\bibinfo {title} {Time-dependent density-matrix
  renormalization-group using adaptive effective hilbert spaces},\ }\href
  {https://doi.org/10.1088/1742-5468/2004/04/P04005} {\bibfield  {journal}
  {\bibinfo  {journal} {Journal of Statistical Mechanics: Theory and
  Experiment}\ }\textbf {\bibinfo {volume} {2004}},\ \bibinfo {pages} {P04005}
  (\bibinfo {year} {2004})}\BibitemShut {NoStop}%
\bibitem [{\citenamefont {Sch\"on}\ \emph {et~al.}(2005)\citenamefont
  {Sch\"on}, \citenamefont {Solano}, \citenamefont {Verstraete}, \citenamefont
  {Cirac},\ and\ \citenamefont {Wolf}}]{PhysRevLett.95.110503}%
  \BibitemOpen
  \bibfield  {author} {\bibinfo {author} {\bibfnamefont {C.}~\bibnamefont
  {Sch\"on}}, \bibinfo {author} {\bibfnamefont {E.}~\bibnamefont {Solano}},
  \bibinfo {author} {\bibfnamefont {F.}~\bibnamefont {Verstraete}}, \bibinfo
  {author} {\bibfnamefont {J.~I.}\ \bibnamefont {Cirac}},\ and\ \bibinfo
  {author} {\bibfnamefont {M.~M.}\ \bibnamefont {Wolf}},\ }\bibfield  {title}
  {\bibinfo {title} {Sequential generation of entangled multiqubit states},\
  }\href {https://doi.org/10.1103/PhysRevLett.95.110503} {\bibfield  {journal}
  {\bibinfo  {journal} {Phys. Rev. Lett.}\ }\textbf {\bibinfo {volume} {95}},\
  \bibinfo {pages} {110503} (\bibinfo {year} {2005})}\BibitemShut {NoStop}%
\bibitem [{\citenamefont {Foss-Feig}\ \emph {et~al.}(2021)\citenamefont
  {Foss-Feig}, \citenamefont {Hayes}, \citenamefont {Dreiling}, \citenamefont
  {Figgatt}, \citenamefont {Gaebler}, \citenamefont {Moses}, \citenamefont
  {Pino},\ and\ \citenamefont {Potter}}]{foss2021holographic}%
  \BibitemOpen
  \bibfield  {author} {\bibinfo {author} {\bibfnamefont {M.}~\bibnamefont
  {Foss-Feig}}, \bibinfo {author} {\bibfnamefont {D.}~\bibnamefont {Hayes}},
  \bibinfo {author} {\bibfnamefont {J.~M.}\ \bibnamefont {Dreiling}}, \bibinfo
  {author} {\bibfnamefont {C.}~\bibnamefont {Figgatt}}, \bibinfo {author}
  {\bibfnamefont {J.~P.}\ \bibnamefont {Gaebler}}, \bibinfo {author}
  {\bibfnamefont {S.~A.}\ \bibnamefont {Moses}}, \bibinfo {author}
  {\bibfnamefont {J.~M.}\ \bibnamefont {Pino}},\ and\ \bibinfo {author}
  {\bibfnamefont {A.~C.}\ \bibnamefont {Potter}},\ }\bibfield  {title}
  {\bibinfo {title} {Holographic quantum algorithms for simulating correlated
  spin systems},\ }\href@noop {} {\bibfield  {journal} {\bibinfo  {journal}
  {Physical Review Research}\ }\textbf {\bibinfo {volume} {3}},\ \bibinfo
  {pages} {033002} (\bibinfo {year} {2021})}\BibitemShut {NoStop}%
\bibitem [{\citenamefont {Haghshenas}\ \emph {et~al.}(2022)\citenamefont
  {Haghshenas}, \citenamefont {Gray}, \citenamefont {Potter},\ and\
  \citenamefont {Chan}}]{haghshenas2022variational}%
  \BibitemOpen
  \bibfield  {author} {\bibinfo {author} {\bibfnamefont {R.}~\bibnamefont
  {Haghshenas}}, \bibinfo {author} {\bibfnamefont {J.}~\bibnamefont {Gray}},
  \bibinfo {author} {\bibfnamefont {A.~C.}\ \bibnamefont {Potter}},\ and\
  \bibinfo {author} {\bibfnamefont {G.~K.-L.}\ \bibnamefont {Chan}},\
  }\bibfield  {title} {\bibinfo {title} {Variational power of quantum circuit
  tensor networks},\ }\href {https://doi.org/10.1103/PhysRevX.12.011047}
  {\bibfield  {journal} {\bibinfo  {journal} {Phys. Rev. X}\ }\textbf {\bibinfo
  {volume} {12}},\ \bibinfo {pages} {011047} (\bibinfo {year}
  {2022})}\BibitemShut {NoStop}%
\bibitem [{\citenamefont {Chen}\ \emph {et~al.}(2018)\citenamefont {Chen},
  \citenamefont {Cheng}, \citenamefont {Xie}, \citenamefont {Wang},\ and\
  \citenamefont {Xiang}}]{PhysRevB.97.085104}%
  \BibitemOpen
  \bibfield  {author} {\bibinfo {author} {\bibfnamefont {J.}~\bibnamefont
  {Chen}}, \bibinfo {author} {\bibfnamefont {S.}~\bibnamefont {Cheng}},
  \bibinfo {author} {\bibfnamefont {H.}~\bibnamefont {Xie}}, \bibinfo {author}
  {\bibfnamefont {L.}~\bibnamefont {Wang}},\ and\ \bibinfo {author}
  {\bibfnamefont {T.}~\bibnamefont {Xiang}},\ }\bibfield  {title} {\bibinfo
  {title} {Equivalence of restricted boltzmann machines and tensor network
  states},\ }\href {https://doi.org/10.1103/PhysRevB.97.085104} {\bibfield
  {journal} {\bibinfo  {journal} {Phys. Rev. B}\ }\textbf {\bibinfo {volume}
  {97}},\ \bibinfo {pages} {085104} (\bibinfo {year} {2018})}\BibitemShut
  {NoStop}%
\bibitem [{\citenamefont {Levine}\ \emph {et~al.}(2019)\citenamefont {Levine},
  \citenamefont {Sharir}, \citenamefont {Cohen},\ and\ \citenamefont
  {Shashua}}]{PhysRevLett.122.065301}%
  \BibitemOpen
  \bibfield  {author} {\bibinfo {author} {\bibfnamefont {Y.}~\bibnamefont
  {Levine}}, \bibinfo {author} {\bibfnamefont {O.}~\bibnamefont {Sharir}},
  \bibinfo {author} {\bibfnamefont {N.}~\bibnamefont {Cohen}},\ and\ \bibinfo
  {author} {\bibfnamefont {A.}~\bibnamefont {Shashua}},\ }\bibfield  {title}
  {\bibinfo {title} {Quantum entanglement in deep learning architectures},\
  }\href {https://doi.org/10.1103/PhysRevLett.122.065301} {\bibfield  {journal}
  {\bibinfo  {journal} {Phys. Rev. Lett.}\ }\textbf {\bibinfo {volume} {122}},\
  \bibinfo {pages} {065301} (\bibinfo {year} {2019})}\BibitemShut {NoStop}%
\bibitem [{\citenamefont {Levine}\ \emph {et~al.}(2017)\citenamefont {Levine},
  \citenamefont {Yakira}, \citenamefont {Cohen},\ and\ \citenamefont
  {Shashua}}]{Levine2017_arxiv}%
  \BibitemOpen
  \bibfield  {author} {\bibinfo {author} {\bibfnamefont {Y.}~\bibnamefont
  {Levine}}, \bibinfo {author} {\bibfnamefont {D.}~\bibnamefont {Yakira}},
  \bibinfo {author} {\bibfnamefont {N.}~\bibnamefont {Cohen}},\ and\ \bibinfo
  {author} {\bibfnamefont {A.}~\bibnamefont {Shashua}},\ }\href
  {https://doi.org/10.48550/ARXIV.1704.01552} {\bibinfo {title} {Deep learning
  and quantum entanglement: Fundamental connections with implications to
  network design}} (\bibinfo {year} {2017})\BibitemShut {NoStop}%
\bibitem [{\citenamefont {Cheng}\ \emph {et~al.}(2019)\citenamefont {Cheng},
  \citenamefont {Wang}, \citenamefont {Xiang},\ and\ \citenamefont
  {Zhang}}]{PhysRevB.99.155131}%
  \BibitemOpen
  \bibfield  {author} {\bibinfo {author} {\bibfnamefont {S.}~\bibnamefont
  {Cheng}}, \bibinfo {author} {\bibfnamefont {L.}~\bibnamefont {Wang}},
  \bibinfo {author} {\bibfnamefont {T.}~\bibnamefont {Xiang}},\ and\ \bibinfo
  {author} {\bibfnamefont {P.}~\bibnamefont {Zhang}},\ }\bibfield  {title}
  {\bibinfo {title} {Tree tensor networks for generative modeling},\ }\href
  {https://doi.org/10.1103/PhysRevB.99.155131} {\bibfield  {journal} {\bibinfo
  {journal} {Phys. Rev. B}\ }\textbf {\bibinfo {volume} {99}},\ \bibinfo
  {pages} {155131} (\bibinfo {year} {2019})}\BibitemShut {NoStop}%
\bibitem [{\citenamefont {Liu}\ \emph {et~al.}(2019)\citenamefont {Liu},
  \citenamefont {Ran}, \citenamefont {Wittek}, \citenamefont {Peng},
  \citenamefont {García}, \citenamefont {Su},\ and\ \citenamefont
  {Lewenstein}}]{Liu_2019}%
  \BibitemOpen
  \bibfield  {author} {\bibinfo {author} {\bibfnamefont {D.}~\bibnamefont
  {Liu}}, \bibinfo {author} {\bibfnamefont {S.-J.}\ \bibnamefont {Ran}},
  \bibinfo {author} {\bibfnamefont {P.}~\bibnamefont {Wittek}}, \bibinfo
  {author} {\bibfnamefont {C.}~\bibnamefont {Peng}}, \bibinfo {author}
  {\bibfnamefont {R.~B.}\ \bibnamefont {García}}, \bibinfo {author}
  {\bibfnamefont {G.}~\bibnamefont {Su}},\ and\ \bibinfo {author}
  {\bibfnamefont {M.}~\bibnamefont {Lewenstein}},\ }\bibfield  {title}
  {\bibinfo {title} {Machine learning by unitary tensor network of hierarchical
  tree structure},\ }\href {https://doi.org/10.1088/1367-2630/ab31ef}
  {\bibfield  {journal} {\bibinfo  {journal} {New Journal of Physics}\ }\textbf
  {\bibinfo {volume} {21}},\ \bibinfo {pages} {073059} (\bibinfo {year}
  {2019})}\BibitemShut {NoStop}%
\bibitem [{\citenamefont {Efthymiou}\ \emph {et~al.}(2019)\citenamefont
  {Efthymiou}, \citenamefont {Hidary},\ and\ \citenamefont
  {Leichenauer}}]{Efthymiou_2019_arxiv}%
  \BibitemOpen
  \bibfield  {author} {\bibinfo {author} {\bibfnamefont {S.}~\bibnamefont
  {Efthymiou}}, \bibinfo {author} {\bibfnamefont {J.}~\bibnamefont {Hidary}},\
  and\ \bibinfo {author} {\bibfnamefont {S.}~\bibnamefont {Leichenauer}},\
  }\href {https://doi.org/10.48550/ARXIV.1906.06329} {\bibinfo {title}
  {Tensornetwork for machine learning}} (\bibinfo {year} {2019})\BibitemShut
  {NoStop}%
\bibitem [{\citenamefont {Wall}\ \emph {et~al.}(2021)\citenamefont {Wall},
  \citenamefont {Abernathy},\ and\ \citenamefont
  {Quiroz}}]{PhysRevResearch.3.023010}%
  \BibitemOpen
  \bibfield  {author} {\bibinfo {author} {\bibfnamefont {M.~L.}\ \bibnamefont
  {Wall}}, \bibinfo {author} {\bibfnamefont {M.~R.}\ \bibnamefont
  {Abernathy}},\ and\ \bibinfo {author} {\bibfnamefont {G.}~\bibnamefont
  {Quiroz}},\ }\bibfield  {title} {\bibinfo {title} {Generative machine
  learning with tensor networks: Benchmarks on near-term quantum computers},\
  }\href {https://doi.org/10.1103/PhysRevResearch.3.023010} {\bibfield
  {journal} {\bibinfo  {journal} {Phys. Rev. Research}\ }\textbf {\bibinfo
  {volume} {3}},\ \bibinfo {pages} {023010} (\bibinfo {year}
  {2021})}\BibitemShut {NoStop}%
\bibitem [{\citenamefont {Han}\ \emph {et~al.}(2018)\citenamefont {Han},
  \citenamefont {Wang}, \citenamefont {Fan}, \citenamefont {Wang},\ and\
  \citenamefont {Zhang}}]{PhysRevX.8.031012}%
  \BibitemOpen
  \bibfield  {author} {\bibinfo {author} {\bibfnamefont {Z.-Y.}\ \bibnamefont
  {Han}}, \bibinfo {author} {\bibfnamefont {J.}~\bibnamefont {Wang}}, \bibinfo
  {author} {\bibfnamefont {H.}~\bibnamefont {Fan}}, \bibinfo {author}
  {\bibfnamefont {L.}~\bibnamefont {Wang}},\ and\ \bibinfo {author}
  {\bibfnamefont {P.}~\bibnamefont {Zhang}},\ }\bibfield  {title} {\bibinfo
  {title} {Unsupervised generative modeling using matrix product states},\
  }\href {https://doi.org/10.1103/PhysRevX.8.031012} {\bibfield  {journal}
  {\bibinfo  {journal} {Phys. Rev. X}\ }\textbf {\bibinfo {volume} {8}},\
  \bibinfo {pages} {031012} (\bibinfo {year} {2018})}\BibitemShut {NoStop}%
\bibitem [{\citenamefont {Cheng}\ \emph {et~al.}(2021)\citenamefont {Cheng},
  \citenamefont {Wang},\ and\ \citenamefont {Zhang}}]{PhysRevB.103.125117}%
  \BibitemOpen
  \bibfield  {author} {\bibinfo {author} {\bibfnamefont {S.}~\bibnamefont
  {Cheng}}, \bibinfo {author} {\bibfnamefont {L.}~\bibnamefont {Wang}},\ and\
  \bibinfo {author} {\bibfnamefont {P.}~\bibnamefont {Zhang}},\ }\bibfield
  {title} {\bibinfo {title} {Supervised learning with projected entangled pair
  states},\ }\href {https://doi.org/10.1103/PhysRevB.103.125117} {\bibfield
  {journal} {\bibinfo  {journal} {Phys. Rev. B}\ }\textbf {\bibinfo {volume}
  {103}},\ \bibinfo {pages} {125117} (\bibinfo {year} {2021})}\BibitemShut
  {NoStop}%
\bibitem [{\citenamefont {Dymarsky}\ and\ \citenamefont
  {Pavlenko}(2022)}]{PhysRevResearch.4.043111}%
  \BibitemOpen
  \bibfield  {author} {\bibinfo {author} {\bibfnamefont {A.}~\bibnamefont
  {Dymarsky}}\ and\ \bibinfo {author} {\bibfnamefont {K.}~\bibnamefont
  {Pavlenko}},\ }\bibfield  {title} {\bibinfo {title} {Tensor network to learn
  the wave function of data},\ }\href
  {https://doi.org/10.1103/PhysRevResearch.4.043111} {\bibfield  {journal}
  {\bibinfo  {journal} {Phys. Rev. Res.}\ }\textbf {\bibinfo {volume} {4}},\
  \bibinfo {pages} {043111} (\bibinfo {year} {2022})}\BibitemShut {NoStop}%
\bibitem [{\citenamefont {Avdoshkin}\ and\ \citenamefont
  {Milekhin}()}]{MPSAdS}%
  \BibitemOpen
  \bibfield  {author} {\bibinfo {author} {\bibnamefont {Avdoshkin}}\ and\
  \bibinfo {author} {\bibnamefont {Milekhin}},\ }\bibfield  {title} {\bibinfo
  {title} {Matrix product states as holographic semi-fixed area states},\
  }\href@noop {} {\bibinfo  {journal} {to appear}\ }\BibitemShut {NoStop}%
\bibitem [{\citenamefont {Kibble}(1976)}]{Kibble1976}%
  \BibitemOpen
\bibfield  {journal} {  }\bibfield  {author} {\bibinfo {author} {\bibfnamefont
  {T.~W.~B.}\ \bibnamefont {Kibble}},\ }\bibfield  {title} {\bibinfo {title}
  {Topology of cosmic domains and strings},\ }\href
  {https://doi.org/10.1088/0305-4470/9/8/029} {\bibfield  {journal} {\bibinfo
  {journal} {Journal of Physics A: Mathematical and General}\ }\textbf
  {\bibinfo {volume} {9}},\ \bibinfo {pages} {1387} (\bibinfo {year}
  {1976})}\BibitemShut {NoStop}%
\bibitem [{\citenamefont {Kibble}(1980)}]{Kibble1980}%
  \BibitemOpen
  \bibfield  {author} {\bibinfo {author} {\bibfnamefont {T.}~\bibnamefont
  {Kibble}},\ }\bibfield  {title} {\bibinfo {title} {Some implications of a
  cosmological phase transition},\ }\href
  {https://doi.org/https://doi.org/10.1016/0370-1573(80)90091-5} {\bibfield
  {journal} {\bibinfo  {journal} {Physics Reports}\ }\textbf {\bibinfo {volume}
  {67}},\ \bibinfo {pages} {183} (\bibinfo {year} {1980})}\BibitemShut
  {NoStop}%
\bibitem [{\citenamefont {Zurek}(1985)}]{zurek1985cosmological}%
  \BibitemOpen
  \bibfield  {author} {\bibinfo {author} {\bibfnamefont {W.~H.}\ \bibnamefont
  {Zurek}},\ }\bibfield  {title} {\bibinfo {title} {Cosmological experiments in
  superfluid helium?},\ }\href@noop {} {\bibfield  {journal} {\bibinfo
  {journal} {Nature}\ }\textbf {\bibinfo {volume} {317}},\ \bibinfo {pages}
  {505} (\bibinfo {year} {1985})}\BibitemShut {NoStop}%
\bibitem [{\citenamefont {del Campo}\ and\ \citenamefont
  {Zurek}(2014)}]{Zurek2014}%
  \BibitemOpen
  \bibfield  {author} {\bibinfo {author} {\bibfnamefont {A.}~\bibnamefont {del
  Campo}}\ and\ \bibinfo {author} {\bibfnamefont {W.~H.}\ \bibnamefont
  {Zurek}},\ }\bibfield  {title} {\bibinfo {title} {Universality of phase
  transition dynamics: Topological defects from symmetry breaking},\ }\href
  {https://www.worldscientific.com/doi/abs/10.1142/S0217751X1430018X}
  {\bibfield  {journal} {\bibinfo  {journal} {Int. J. Mod. Phys. A}\ }\textbf
  {\bibinfo {volume} {29}},\ \bibinfo {pages} {1430018} (\bibinfo {year}
  {2014})}\BibitemShut {NoStop}%
\bibitem [{\citenamefont {Sachdev}(1999)}]{sachdev1999quantum}%
  \BibitemOpen
  \bibfield  {author} {\bibinfo {author} {\bibfnamefont {S.}~\bibnamefont
  {Sachdev}},\ }\bibfield  {title} {\bibinfo {title} {Quantum phase
  transitions},\ }\href@noop {} {\bibfield  {journal} {\bibinfo  {journal}
  {Physics world}\ }\textbf {\bibinfo {volume} {12}},\ \bibinfo {pages} {33}
  (\bibinfo {year} {1999})}\BibitemShut {NoStop}%
\bibitem [{\citenamefont {De~Grandi}\ and\ \citenamefont
  {Polkovnikov}(2010)}]{DeGrandi2010}%
  \BibitemOpen
  \bibfield  {author} {\bibinfo {author} {\bibfnamefont {C.}~\bibnamefont
  {De~Grandi}}\ and\ \bibinfo {author} {\bibfnamefont {A.}~\bibnamefont
  {Polkovnikov}},\ }\bibinfo {title} {Adiabatic perturbation theory: From
  landau--zener problem to quenching through a quantum critical point},\ in\
  \href {https://doi.org/10.1007/978-3-642-11470-0_4} {\emph {\bibinfo
  {booktitle} {Quantum Quenching, Annealing and Computation}}},\ \bibinfo
  {editor} {edited by\ \bibinfo {editor} {\bibfnamefont {A.~K.}\ \bibnamefont
  {Chandra}}, \bibinfo {editor} {\bibfnamefont {A.}~\bibnamefont {Das}},\ and\
  \bibinfo {editor} {\bibfnamefont {B.~K.}\ \bibnamefont {Chakrabarti}}}\
  (\bibinfo  {publisher} {Springer Berlin Heidelberg},\ \bibinfo {address}
  {Berlin, Heidelberg},\ \bibinfo {year} {2010})\ pp.\ \bibinfo {pages}
  {75--114}\BibitemShut {NoStop}%
\bibitem [{\citenamefont {Rossini}\ and\ \citenamefont
  {Vicari}(2021)}]{ROSSINI20211}%
  \BibitemOpen
  \bibfield  {author} {\bibinfo {author} {\bibfnamefont {D.}~\bibnamefont
  {Rossini}}\ and\ \bibinfo {author} {\bibfnamefont {E.}~\bibnamefont
  {Vicari}},\ }\bibfield  {title} {\bibinfo {title} {Coherent and dissipative
  dynamics at quantum phase transitions},\ }\href
  {https://doi.org/https://doi.org/10.1016/j.physrep.2021.08.003} {\bibfield
  {journal} {\bibinfo  {journal} {Physics Reports}\ }\textbf {\bibinfo {volume}
  {936}},\ \bibinfo {pages} {1} (\bibinfo {year} {2021})},\ \bibinfo {note}
  {coherent and dissipative dynamics at quantum phase transitions}\BibitemShut
  {NoStop}%
\bibitem [{\citenamefont {Clark}\ \emph {et~al.}(2016)\citenamefont {Clark},
  \citenamefont {Feng},\ and\ \citenamefont {Chin}}]{Clark_2016}%
  \BibitemOpen
  \bibfield  {author} {\bibinfo {author} {\bibfnamefont {L.~W.}\ \bibnamefont
  {Clark}}, \bibinfo {author} {\bibfnamefont {L.}~\bibnamefont {Feng}},\ and\
  \bibinfo {author} {\bibfnamefont {C.}~\bibnamefont {Chin}},\ }\bibfield
  {title} {\bibinfo {title} {Universal space-time scaling symmetry in the
  dynamics of bosons across a quantum phase transition},\ }\href
  {https://doi.org/10.1126/science.aaf9657} {\bibfield  {journal} {\bibinfo
  {journal} {Science}\ }\textbf {\bibinfo {volume} {354}},\ \bibinfo {pages}
  {606} (\bibinfo {year} {2016})},\ \Eprint
  {https://arxiv.org/abs/https://www.science.org/doi/pdf/10.1126/science.aaf9657}
  {https://www.science.org/doi/pdf/10.1126/science.aaf9657} \BibitemShut
  {NoStop}%
\bibitem [{\citenamefont {Dziarmaga}(2005)}]{dziarmaga2005dynamics}%
  \BibitemOpen
  \bibfield  {author} {\bibinfo {author} {\bibfnamefont {J.}~\bibnamefont
  {Dziarmaga}},\ }\bibfield  {title} {\bibinfo {title} {Dynamics of a quantum
  phase transition: Exact solution of the quantum ising model},\ }\href
  {https://doi.org/10.1103/PhysRevLett.95.245701} {\bibfield  {journal}
  {\bibinfo  {journal} {Phys. Rev. Lett.}\ }\textbf {\bibinfo {volume} {95}},\
  \bibinfo {pages} {245701} (\bibinfo {year} {2005})}\BibitemShut {NoStop}%
\bibitem [{\citenamefont {Cherng}\ and\ \citenamefont
  {Levitov}(2006)}]{PhysRevA.73.043614}%
  \BibitemOpen
  \bibfield  {author} {\bibinfo {author} {\bibfnamefont {R.~W.}\ \bibnamefont
  {Cherng}}\ and\ \bibinfo {author} {\bibfnamefont {L.~S.}\ \bibnamefont
  {Levitov}},\ }\bibfield  {title} {\bibinfo {title} {Entropy and correlation
  functions of a driven quantum spin chain},\ }\href
  {https://doi.org/10.1103/PhysRevA.73.043614} {\bibfield  {journal} {\bibinfo
  {journal} {Phys. Rev. A}\ }\textbf {\bibinfo {volume} {73}},\ \bibinfo
  {pages} {043614} (\bibinfo {year} {2006})}\BibitemShut {NoStop}%
\bibitem [{\citenamefont {De~Grandi}\ \emph {et~al.}(2011)\citenamefont
  {De~Grandi}, \citenamefont {Polkovnikov},\ and\ \citenamefont
  {Sandvik}}]{PhysRevB.84.224303}%
  \BibitemOpen
  \bibfield  {author} {\bibinfo {author} {\bibfnamefont {C.}~\bibnamefont
  {De~Grandi}}, \bibinfo {author} {\bibfnamefont {A.}~\bibnamefont
  {Polkovnikov}},\ and\ \bibinfo {author} {\bibfnamefont {A.~W.}\ \bibnamefont
  {Sandvik}},\ }\bibfield  {title} {\bibinfo {title} {Universal nonequilibrium
  quantum dynamics in imaginary time},\ }\href
  {https://doi.org/10.1103/PhysRevB.84.224303} {\bibfield  {journal} {\bibinfo
  {journal} {Phys. Rev. B}\ }\textbf {\bibinfo {volume} {84}},\ \bibinfo
  {pages} {224303} (\bibinfo {year} {2011})}\BibitemShut {NoStop}%
\bibitem [{\citenamefont {Calabrese}\ and\ \citenamefont
  {Cardy}(2009)}]{Calabrese_2009}%
  \BibitemOpen
  \bibfield  {author} {\bibinfo {author} {\bibfnamefont {P.}~\bibnamefont
  {Calabrese}}\ and\ \bibinfo {author} {\bibfnamefont {J.}~\bibnamefont
  {Cardy}},\ }\bibfield  {title} {\bibinfo {title} {Entanglement entropy and
  conformal field theory},\ }\href
  {https://doi.org/10.1088/1751-8113/42/50/504005} {\bibfield  {journal}
  {\bibinfo  {journal} {Journal of Physics A: Mathematical and Theoretical}\
  }\textbf {\bibinfo {volume} {42}},\ \bibinfo {pages} {504005} (\bibinfo
  {year} {2009})}\BibitemShut {NoStop}%
\bibitem [{\citenamefont {Cao}\ \emph {et~al.}(2018)\citenamefont {Cao},
  \citenamefont {Hu},\ and\ \citenamefont {Zhong}}]{PhysRevB.98.245124}%
  \BibitemOpen
  \bibfield  {author} {\bibinfo {author} {\bibfnamefont {X.}~\bibnamefont
  {Cao}}, \bibinfo {author} {\bibfnamefont {Q.}~\bibnamefont {Hu}},\ and\
  \bibinfo {author} {\bibfnamefont {F.}~\bibnamefont {Zhong}},\ }\bibfield
  {title} {\bibinfo {title} {Scaling theory of entanglement entropy in
  confinements near quantum critical points},\ }\href
  {https://doi.org/10.1103/PhysRevB.98.245124} {\bibfield  {journal} {\bibinfo
  {journal} {Phys. Rev. B}\ }\textbf {\bibinfo {volume} {98}},\ \bibinfo
  {pages} {245124} (\bibinfo {year} {2018})}\BibitemShut {NoStop}%
\bibitem [{\citenamefont {Tagliacozzo}\ \emph {et~al.}(2008)\citenamefont
  {Tagliacozzo}, \citenamefont {de~Oliveira}, \citenamefont {Iblisdir},\ and\
  \citenamefont {Latorre}}]{tagliacozzo2008scaling}%
  \BibitemOpen
  \bibfield  {author} {\bibinfo {author} {\bibfnamefont {L.}~\bibnamefont
  {Tagliacozzo}}, \bibinfo {author} {\bibfnamefont {T.~R.}\ \bibnamefont
  {de~Oliveira}}, \bibinfo {author} {\bibfnamefont {S.}~\bibnamefont
  {Iblisdir}},\ and\ \bibinfo {author} {\bibfnamefont {J.~I.}\ \bibnamefont
  {Latorre}},\ }\bibfield  {title} {\bibinfo {title} {Scaling of entanglement
  support for matrix product states},\ }\href
  {https://doi.org/10.1103/PhysRevB.78.024410} {\bibfield  {journal} {\bibinfo
  {journal} {Phys. Rev. B}\ }\textbf {\bibinfo {volume} {78}},\ \bibinfo
  {pages} {024410} (\bibinfo {year} {2008})}\BibitemShut {NoStop}%
\bibitem [{\citenamefont {Pollmann}\ \emph {et~al.}(2009)\citenamefont
  {Pollmann}, \citenamefont {Mukerjee}, \citenamefont {Turner},\ and\
  \citenamefont {Moore}}]{pollmann2009theory}%
  \BibitemOpen
  \bibfield  {author} {\bibinfo {author} {\bibfnamefont {F.}~\bibnamefont
  {Pollmann}}, \bibinfo {author} {\bibfnamefont {S.}~\bibnamefont {Mukerjee}},
  \bibinfo {author} {\bibfnamefont {A.~M.}\ \bibnamefont {Turner}},\ and\
  \bibinfo {author} {\bibfnamefont {J.~E.}\ \bibnamefont {Moore}},\ }\bibfield
  {title} {\bibinfo {title} {Theory of finite-entanglement scaling at
  one-dimensional quantum critical points},\ }\href
  {https://doi.org/10.1103/PhysRevLett.102.255701} {\bibfield  {journal}
  {\bibinfo  {journal} {Phys. Rev. Lett.}\ }\textbf {\bibinfo {volume} {102}},\
  \bibinfo {pages} {255701} (\bibinfo {year} {2009})}\BibitemShut {NoStop}%
\bibitem [{\citenamefont {Pirvu}\ \emph {et~al.}(2012)\citenamefont {Pirvu},
  \citenamefont {Vidal}, \citenamefont {Verstraete},\ and\ \citenamefont
  {Tagliacozzo}}]{pirvu2012}%
  \BibitemOpen
  \bibfield  {author} {\bibinfo {author} {\bibfnamefont {B.}~\bibnamefont
  {Pirvu}}, \bibinfo {author} {\bibfnamefont {G.}~\bibnamefont {Vidal}},
  \bibinfo {author} {\bibfnamefont {F.}~\bibnamefont {Verstraete}},\ and\
  \bibinfo {author} {\bibfnamefont {L.}~\bibnamefont {Tagliacozzo}},\
  }\bibfield  {title} {\bibinfo {title} {Matrix product states for critical
  spin chains: Finite-size versus finite-entanglement scaling},\ }\href
  {https://doi.org/10.1103/PhysRevB.86.075117} {\bibfield  {journal} {\bibinfo
  {journal} {Phys. Rev. B}\ }\textbf {\bibinfo {volume} {86}},\ \bibinfo
  {pages} {075117} (\bibinfo {year} {2012})}\BibitemShut {NoStop}%
\bibitem [{\citenamefont {Francesco}\ \emph {et~al.}(2012)\citenamefont
  {Francesco}, \citenamefont {Mathieu},\ and\ \citenamefont
  {S{\'e}n{\'e}chal}}]{francesco2012conformal}%
  \BibitemOpen
  \bibfield  {author} {\bibinfo {author} {\bibfnamefont {P.}~\bibnamefont
  {Francesco}}, \bibinfo {author} {\bibfnamefont {P.}~\bibnamefont {Mathieu}},\
  and\ \bibinfo {author} {\bibfnamefont {D.}~\bibnamefont {S{\'e}n{\'e}chal}},\
  }\href@noop {} {\emph {\bibinfo {title} {Conformal field theory}}}\ (\bibinfo
   {publisher} {Springer Science \& Business Media},\ \bibinfo {year}
  {2012})\BibitemShut {NoStop}%
\bibitem [{\citenamefont {Barthel}\ and\ \citenamefont
  {Zhang}(2020)}]{Barthel_2020}%
  \BibitemOpen
  \bibfield  {author} {\bibinfo {author} {\bibfnamefont {T.}~\bibnamefont
  {Barthel}}\ and\ \bibinfo {author} {\bibfnamefont {Y.}~\bibnamefont
  {Zhang}},\ }\bibfield  {title} {\bibinfo {title} {Optimized
  lie{\textendash}trotter{\textendash}suzuki decompositions for two and three
  non-commuting terms},\ }\href {https://doi.org/10.1016/j.aop.2020.168165}
  {\bibfield  {journal} {\bibinfo  {journal} {Annals of Physics}\ }\textbf
  {\bibinfo {volume} {418}},\ \bibinfo {pages} {168165} (\bibinfo {year}
  {2020})}\BibitemShut {NoStop}%
\bibitem [{\citenamefont {S\'olyom}\ and\ \citenamefont
  {Pfeuty}(1981)}]{Solyom_1981}%
  \BibitemOpen
  \bibfield  {author} {\bibinfo {author} {\bibfnamefont {J.}~\bibnamefont
  {S\'olyom}}\ and\ \bibinfo {author} {\bibfnamefont {P.}~\bibnamefont
  {Pfeuty}},\ }\bibfield  {title} {\bibinfo {title} {Renormalization-group
  study of the hamiltonian version of the potts model},\ }\href
  {https://doi.org/10.1103/PhysRevB.24.218} {\bibfield  {journal} {\bibinfo
  {journal} {Phys. Rev. B}\ }\textbf {\bibinfo {volume} {24}},\ \bibinfo
  {pages} {218} (\bibinfo {year} {1981})}\BibitemShut {NoStop}%
\bibitem [{\citenamefont {Gong}\ \emph {et~al.}(2010)\citenamefont {Gong},
  \citenamefont {Zhong}, \citenamefont {Huang},\ and\ \citenamefont
  {Fan}}]{Gong_2010}%
  \BibitemOpen
  \bibfield  {author} {\bibinfo {author} {\bibfnamefont {S.}~\bibnamefont
  {Gong}}, \bibinfo {author} {\bibfnamefont {F.}~\bibnamefont {Zhong}},
  \bibinfo {author} {\bibfnamefont {X.}~\bibnamefont {Huang}},\ and\ \bibinfo
  {author} {\bibfnamefont {S.}~\bibnamefont {Fan}},\ }\bibfield  {title}
  {\bibinfo {title} {Finite-time scaling via linear driving},\ }\href
  {https://doi.org/10.1088/1367-2630/12/4/043036} {\bibfield  {journal}
  {\bibinfo  {journal} {New Journal of Physics}\ }\textbf {\bibinfo {volume}
  {12}},\ \bibinfo {pages} {043036} (\bibinfo {year} {2010})}\BibitemShut
  {NoStop}%
\bibitem [{\citenamefont {Huang}\ \emph {et~al.}(2014)\citenamefont {Huang},
  \citenamefont {Yin}, \citenamefont {Feng},\ and\ \citenamefont
  {Zhong}}]{PhysRevB.90.134108}%
  \BibitemOpen
  \bibfield  {author} {\bibinfo {author} {\bibfnamefont {Y.}~\bibnamefont
  {Huang}}, \bibinfo {author} {\bibfnamefont {S.}~\bibnamefont {Yin}}, \bibinfo
  {author} {\bibfnamefont {B.}~\bibnamefont {Feng}},\ and\ \bibinfo {author}
  {\bibfnamefont {F.}~\bibnamefont {Zhong}},\ }\bibfield  {title} {\bibinfo
  {title} {Kibble-zurek mechanism and finite-time scaling},\ }\href
  {https://doi.org/10.1103/PhysRevB.90.134108} {\bibfield  {journal} {\bibinfo
  {journal} {Phys. Rev. B}\ }\textbf {\bibinfo {volume} {90}},\ \bibinfo
  {pages} {134108} (\bibinfo {year} {2014})}\BibitemShut {NoStop}%
\bibitem [{\citenamefont {Vanderstraeten}\ \emph
  {et~al.}(2019{\natexlab{b}})\citenamefont {Vanderstraeten}, \citenamefont
  {Haegeman},\ and\ \citenamefont {Verstraete}}]{vanderstraeten2019tangent}%
  \BibitemOpen
  \bibfield  {author} {\bibinfo {author} {\bibfnamefont {L.}~\bibnamefont
  {Vanderstraeten}}, \bibinfo {author} {\bibfnamefont {J.}~\bibnamefont
  {Haegeman}},\ and\ \bibinfo {author} {\bibfnamefont {F.}~\bibnamefont
  {Verstraete}},\ }\bibfield  {title} {\bibinfo {title} {Tangent-space methods
  for uniform matrix product states},\ }\href@noop {} {\bibfield  {journal}
  {\bibinfo  {journal} {SciPost Physics Lecture Notes}\ ,\ \bibinfo {pages}
  {007}} (\bibinfo {year} {2019}{\natexlab{b}})}\BibitemShut {NoStop}%
\bibitem [{\citenamefont {Vidal}(2004)}]{vidal2004efficient}%
  \BibitemOpen
  \bibfield  {author} {\bibinfo {author} {\bibfnamefont {G.}~\bibnamefont
  {Vidal}},\ }\bibfield  {title} {\bibinfo {title} {Efficient simulation of
  one-dimensional quantum many-body systems},\ }\href
  {https://doi.org/10.1103/PhysRevLett.93.040502} {\bibfield  {journal}
  {\bibinfo  {journal} {Phys. Rev. Lett.}\ }\textbf {\bibinfo {volume} {93}},\
  \bibinfo {pages} {040502} (\bibinfo {year} {2004})}\BibitemShut {NoStop}%
\bibitem [{\citenamefont {Verstraete}\ \emph {et~al.}(2004)\citenamefont
  {Verstraete}, \citenamefont {Garc\'{\i}a-Ripoll},\ and\ \citenamefont
  {Cirac}}]{PhysRevLett.93.207204}%
  \BibitemOpen
  \bibfield  {author} {\bibinfo {author} {\bibfnamefont {F.}~\bibnamefont
  {Verstraete}}, \bibinfo {author} {\bibfnamefont {J.~J.}\ \bibnamefont
  {Garc\'{\i}a-Ripoll}},\ and\ \bibinfo {author} {\bibfnamefont {J.~I.}\
  \bibnamefont {Cirac}},\ }\bibfield  {title} {\bibinfo {title} {Matrix product
  density operators: Simulation of finite-temperature and dissipative
  systems},\ }\href {https://doi.org/10.1103/PhysRevLett.93.207204} {\bibfield
  {journal} {\bibinfo  {journal} {Phys. Rev. Lett.}\ }\textbf {\bibinfo
  {volume} {93}},\ \bibinfo {pages} {207204} (\bibinfo {year}
  {2004})}\BibitemShut {NoStop}%
\end{thebibliography}%
\end{document}